\documentclass[aps,prl,reprint,showpacs,superscriptaddress,notitlepage,twocolumn,10pt]{revtex4-2}%
\usepackage{amsfonts}
\usepackage{mathrsfs}
\usepackage{amsmath}
\usepackage{amssymb}
\usepackage{graphicx}
\usepackage{color}%
\usepackage{mathtools}
\usepackage{booktabs}
\usepackage[colorlinks,linkcolor=blue,citecolor=blue,hyperindex,bookmarks=false,pdfstartview=FitH]{hyperref}

\providecommand{\U}[1]{\protect\rule{.1in}{.1in}}

\newcommand{\figpanel}[2]{Fig.~\hyperref[#1]{\ref*{#1}(#2)}}
\newcommand{\figpanels}[3]{Fig.~\hyperref[#1]{\ref*{#1}(#2)-(#3)}}
\newcommand{\figpanelNoPrefix}[2]{\hyperref[#1]{\ref*{#1}(#2)}}

\usepackage{mleftright} 

\begin{document}
\title{Giant Emitters in a Structured Bath with Non-Hermitian Skin Effect}

\author{Lei Du}
\email{lei.du@chalmers.se}
\affiliation{School of Physics and Center for Quantum Sciences, Northeast Normal University, Changchun 130024, China}
\affiliation{Department of Microtechnology and Nanoscience, Chalmers University of Technology, 412 96 Gothenburg, Sweden}
\author{Lingzhen Guo}
\affiliation{Center for Joint Quantum Studies and Department of Physics, School of Science, Tianjin University, Tianjin 300072, China}
\author{Yan Zhang}
\email{zhangy345@nenu.edu.cn}
\affiliation{School of Physics and Center for Quantum Sciences, Northeast Normal University, Changchun 130024, China}
\author{Anton Frisk Kockum}
\affiliation{Department of Microtechnology and Nanoscience, Chalmers University of Technology, 412 96 Gothenburg, Sweden}

\date{\today }

\begin{abstract}
Giant emitters derive their name from nonlocal field-emitter interactions and feature diverse self-interference effects. Most of the existing works on giant emitters have considered Hermitian waveguides or photonic lattices. In this work, we unveil how giant emitters behave if they are coupled to a non-Hermitian bath, i.e., a Hatano-Nelson (HN) model which features a non-Hermitian skin effect due to the asymmetric inter-site tunneling rates. We show that the behaviors of the giant emitters are closely related to the stability of the bath. In the convectively unstable regime, where the HN model can be mapped to a pseudo-Hermitian lattice, a giant emitter can either behave as in a Hermitian bath or undergo excitation amplification, depending on the relative strength of different emitter-bath coupling paths. Based on this mechanism, we can realize protected nonreciprocal interactions between giant emitters, with nonreciprocity opposite to that of the bath. Such giant-emitter effects are not allowed, however, if the HN model enters the absolutely unstable regime, where the coupled emitters always show secular energy growth. Our proposal provides a new paradigm of non-Hermitian quantum optics, which may be useful for, e.g., engineering effective interactions between quantum emitters and performing many-body simulations in the non-Hermitian framework.
\end{abstract}  

\maketitle



\emph{Introduction.}---Giant emitters, which feature (discrete) nonlocal interactions with a bath, are setting up a new quantum optical paradigm and attracting increasing interest~\cite{fiveyear}. A hallmark of giant emitters is that their effective relaxation rates and transition frequencies are closely related to the interference effects of the nonlocal couplings~\cite{LambAFK,SAW2014,GLZ2017,nonexp,WilsonPRA2021}. For example, consider a two-level giant emitter (with transition frequency $\omega_0$) that is coupled to a one-dimensional waveguide at two points $x=0$ and $x=d$ with identical coupling strength $g$. Its effective relaxation rate (to the waveguide) is given by~\cite{LambAFK,braided}
\begin{equation}
\Gamma_{\text{eff}} = \text{Re} \mleft[ 4\pi g^{2} J(\omega_{0}) \mleft( 1 + e^{ik_{0}d} \mright) \mright],
\label{eq1}
\end{equation}
where $k_{0}$ and $J(\omega_{0})$ are the wave number and the density of states of the waveguide field at frequency $\omega_{0}$, respectively. Clearly, the relaxation of the giant emitter is inhibited (enhanced) if $k_{0}d$ is an odd (even) multiple of $\pi$, which can be understood as the destructive (constructive) interference of the two coupling paths. Based on such interference effects, one can realize decoherence-free interaction (DFI) between giant emitters via the waveguide, provided that they are coupled to the waveguide in a ``braided'' structure with interleaved coupling points~\cite{NoriGA,braided,FCdeco,AFKchiral,complexDFI}. This DFI is essentially different from DFIs arising from the overlaps of photon-atom bound states~\cite{SDFI1,SDFI2,SDFI3}, which require that the frequencies of the emitters fall within the band gap of the (structured) bath. Moreover, it is also possible to realize chiral spontaneous emission and bound states if an additional phase difference is encoded into the nonlocal emitter-bath interaction~\cite{WXchiral1,WXchiral2,DLprl,CYTcp,LeiQST,MohammadArxiv}.

The above giant-emitter effects are based on (structured) waveguides governed by Bloch's theorem~\cite{Bloch}, which states that the wave functions of a translationally invariant Hermitian system are plane waves modulated by a spatially periodic phase factor. For the giant-emitter system above, this is manifested by the waveguide field acquiring a phase $k_{0}d$ when traveling between coupling points. For clarity, hereafter we refer to the above effects as ``conventional giant-emitter interference effects'' (CGIEs). In contrast, non-Hermitian systems exhibit a plethora of peculiar properties with no Hermitian analogues~\cite{NHg1,NHg2,NHg3,NHg4,NHg5,NHg6,NHg7,GongPRX,NHg8}, such as exceptional points~\cite{EpHeiss,LiouvillianEp,EpAlu,EpTopo} and biorthogonal eigenstates~\cite{biortho1,biortho2}. In particular, the conventional Bloch theorem can break down in a class of non-Hermitian systems, which feature non-Hermitian skin effects and are governed by non-Bloch band theory~\cite{NB1,NB2,NB3,NB4,NB5}. The most typical example is the Hatano-Nelson (HN) model~\cite{HNmodel}, a one-dimensional tight-binding lattice with asymmetric inter-site tunneling rates, whose experimental implementations include coupled-(ring)-resonator arrays with auxiliary couplers with engineered gain and loss~\cite{LonghiSR,Longhi2015prb}, cascaded quantum systems based on reservoir engineering~\cite{AnjaPRX}, discrete-time non-Hermitian quantum walks~\cite{Qwalk1,Qwalk2,Qwalk3}, and photonic synthetic dimensions~\cite{LuSkin2D,FanScience21}. The spectra of such non-Hermitian systems are quite sensitive to the boundary conditions, which can result in unconventional topological~\cite{EpTopo,Yao2018,Yao20182,TopoOrigin,YFChen2020,TopoSensor,HigherOrderSkin} and quantum optical~\cite{LonghiQD,Federico,GongHN,GongHN2} phenomena. In view of this, it is natural to ask \emph{how giant emitters behave in a bath with non-Hermitian skin effect.}


To address this question, we here study the quantum dynamics of giant emitters coupled to an HN model. We reveal that the giant emitters can exhibit essentially different behaviors, depending on what regime of the HN model we consider and the relative strengths of the nonlocal couplings. In particular, we unveil a series of unconventional quantum optical phenomena, such as secular energy growth and protected nonreciprocal interatomic interactions with nonreciprocity opposite to the field, which have no counterparts in small-emitter nor Hermitian giant-emitter systems. These findings may find applications in quantum simulation and inspire further studies in non-Hermitian quantum optics.

\begin{figure}
\centering
\includegraphics[width=\linewidth]{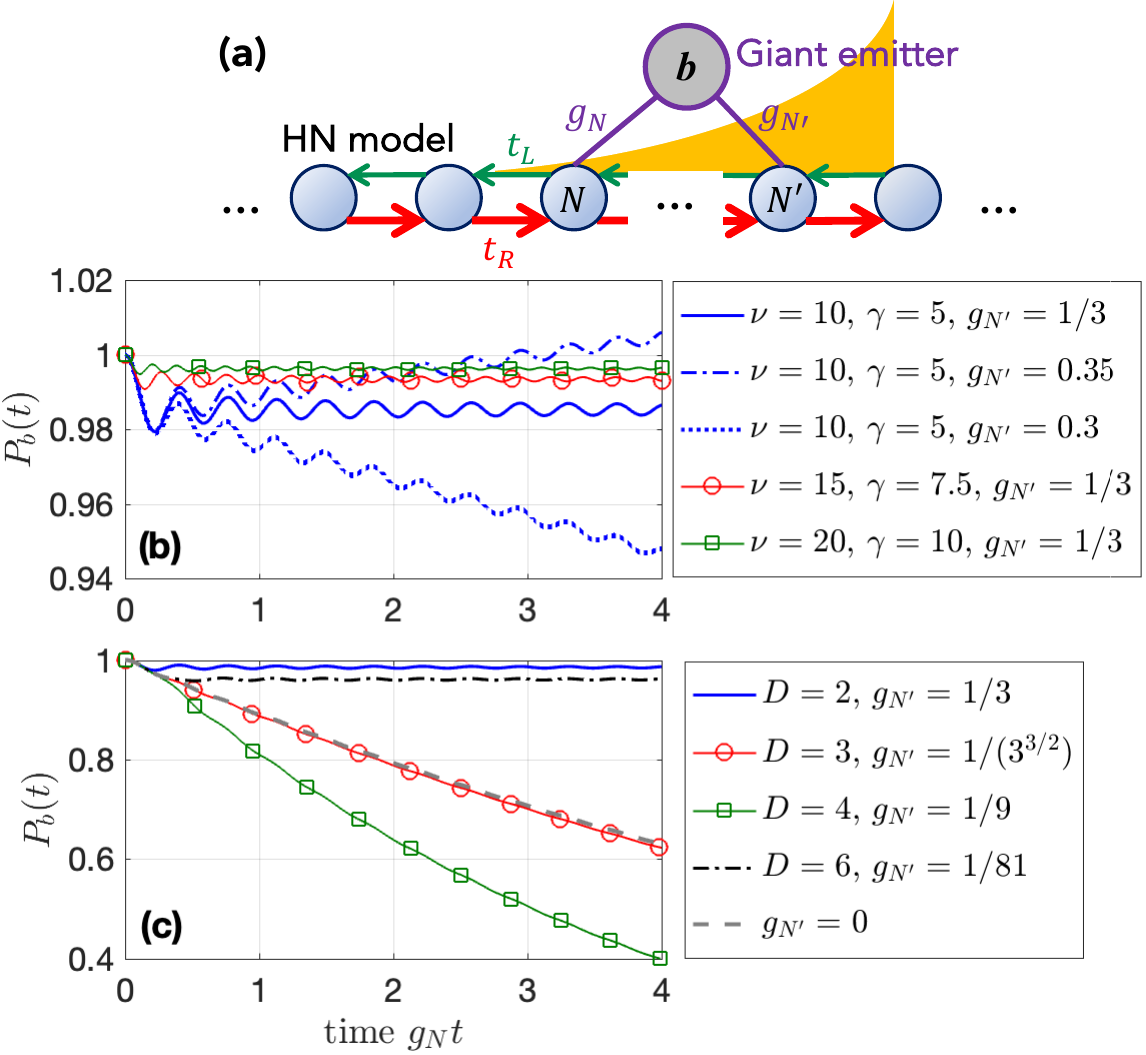}
\caption{(a) Schematic of a giant emitter coupled to an HN model. (b), (c) Time evolution of $P_{b}(t)$ for different parameters. We assume $D=2$ in (b) and $\nu=10$ and $\gamma=5$ in (c). Other parameters are $g_{N}=1$ and $M=1000$.}
\label{Fbasic}
\end{figure}

\emph{Model.}---We first consider a bosonic mode $b$ coupled to two lattice sites of an HN model [\figpanel{Fbasic}{a}]. In the interaction picture, the Hamiltonian is ($\hbar=1$ hereafter)
\begin{equation}
\begin{split}
H&=\sum_{n=0}^{M-1} \mleft( t_{R} a_{n+1}^{\dag}a_{n} + t_{L} a_{n}^{\dag} a_{n+1} \mright)\\
&\quad\, + \mleft( g_{N} b^{\dag} a_{N} + g_{N'} b^{\dag} a_{N'} + \text{H.c.} \mright),
\end{split}
\label{basicH}
\end{equation}
where $a_{n}$ is the annihilation operator of the $n$th site of the HN model (we consider $M$ lattice sites in total; $n\in[0,M-1]$); $t_{R}=\nu+\gamma$ and $t_{L}=\nu-\gamma$ are the nearest-neighbor tunneling amplitudes towards the left and right, respectively, with $\nu$ the base tunneling amplitude and $\gamma$ describing the non-Hermiticity of the system ($H$ becomes Hermitian when $\gamma=0$); $g_{N}$ ($g_{N'}$) is the coupling strength between mode $b$ and the $N$th ($N'$th) site of the HN model. We have assumed that the emitter (i.e., mode $b$) is resonant with the lattice band center. Under open boundary conditions, the wave functions of the bare HN model can be written as $\psi_{n}=\beta^{n}\sin{(k_{l}n)}/\sqrt{(M-1)/2}$ with $\beta=\sqrt{t_{R}/t_{L}}$ and $k_{l}=l\pi/(M-1)$ ($l=1, 2, \ldots, M-2$)~\cite{curveLv}. An imaginary gauge field~\cite{Longhi2015prb}, described by the imaginary wave number $k_{\text{Im}}=-i\text{ln}(\beta)$, modifies the wave functions so that all of them pile up at one of the two lattice edges, depending on the relative strength of $t_{R}$ and $t_{L}$. In other words, fields are amplified (attenuated) when traveling in the direction of the stronger (weaker) tunneling amplitude.


Before proceeding, we briefly motivate choosing a bosonic mode as a generalized giant emitter~\cite{YouNC}. In the non-Hermitian case here, the dynamics do not conserve the norm of the initial state. For a single-excitation initial state, the total excitation number of the system at time $t>0$ can be larger than one, which in a photonic implementation of our model (e.g., a harmonic oscillator coupled to an unstable coupled-resonator array) can be understood as amplified optical intensity arising from the field amplification. In this case, it is more convenient to consider a bosonic mode since the dynamics of a two-level system could exhibit complicated nonlinear characteristics. Alternatively, the single-excitation assumption can be restored by, e.g., introducing uniform on-site losses to the lattice sites~\cite{Federico,GongHN,GongHN2}, as will be discussed below.




\emph{CGIEs in an HN model.}---Now we show that, even in such a non-Bloch bath, it is possible to recover CGIEs by matching the relative coupling strength at the two coupling points with the non-Bloch phase factor $\beta^{n}$. Figures~\figpanelNoPrefix{Fbasic}{b} and \figpanelNoPrefix{Fbasic}{c} show the evolution of the mean particle number $P_{b}(t)=|u_{b}(t)|^{2}$ of the giant emitter $b$, with system state $|\psi_{1}(t)\rangle=[u_{b}(t)b^{\dag}+\sum_{n}u_{a,n}(t)a_{n}^{\dag}]|\emptyset\rangle$ ($|\emptyset\rangle$ is the ground state of the whole model) and initial condition $u_{\alpha}(0)=\delta_{\alpha,b}$~\cite{LonghiQD}. To avoid boundary effects, we consider a long enough HN lattice with $M=1000$ and $0\ll\{N,N'\}\ll M-1$. Moreover, we assume $\{|t_{R}|,|t_{L}|\}\gg |g|$ so that the non-Markovian retardation effect is weak~\cite{DLprl,LonghiGA,DLprr2,AFKstructured}.

Figure \figpanelNoPrefix{Fbasic}{b} shows that the giant emitter exhibits a fractional decay [i.e., $0<P_{b}(t\rightarrow+\infty)<1$] when $g_{N'}/g_{N}=\beta^{-D}$ and $D=N'-N=2$, although it is coupled to a non-Hermitian bath featuring directional field amplification and attenuation. This signifies the recovering of CGIEs in the sense that the emitter can be effectively decoupled from the bath due to the nonlocal interaction. The emitter tends to be completely dissipationless with the increase of the tunneling amplitudes, which is also consistent with the Hermitian case. Note that the fractional decay is sensitive to the relative strength $g_{N'}/g_{N}$: $P_{b}(t)$ will increase (decrease) with time if $g_{N'}/g_{N}$ is larger (smaller) than the critical value $\beta^{-2}$ (cf.~the blue dot-dashed and dotted lines). This effect means that the giant emitter can serve as a precise probe for the non-Hermiticity $\gamma$ of the bath.

We also plot the evolution of $P_{b}(t)$ for different values of $D$ in \figpanel{Fbasic}{c}, with the matching condition $g_{N'}/g_{N}=\beta^{-D}$ always fulfilled. Again, similar to the Hermitian case, the giant emitter exhibits $D$-dependent relaxation dynamics ranging from decoherence-free behavior (i.e., fractional decay for $D = 2, 6$) to superradiance-like behavior [i.e., enhanced relaxation for $D=4$, compared to the small-atom case (gray dashed line)].

The recovered CGIEs can be understood from the self-energy of the giant emitter, which can be obtained (using the resolvent method~\cite{resolvent}; see Sec.~I of~\cite{SM} for details) as
\begin{equation}
\begin{split}
\Sigma_{b}(z)&=\mp \frac{1}{\sqrt{z^{2}-4t_{R}t_{L}}} \mleft[ g_{N}^{2} + g_{N'}^{2} \mright.\\
&\mleft.\quad\,+ g_{N} g_{N'} y_{\pm}^{D} \mleft( \beta^{D} + \beta^{-D} \mright) \mright]
\end{split}
\label{selfE}
\end{equation}
with $y_{\pm}=(z\pm\sqrt{z^{2}-4t_{R}t_{L}})/(2\sqrt{t_{R}t_{L}})$. The real and imaginary parts of $\Sigma_{b}$ represent the Lamb shift and effective relaxation rate, respectively, of $b$ due to its interaction with the bath. Under the Weisskopf-Wigner approximation~\cite{WWapprox} (valid for weak emitter-bath couplings), the dynamics of the emitter are well captured by the self-energy close to the real axis~\cite{AGT2017prl,AGT2017pra}, i.e., $\Sigma_{b}(\Delta+i0^{+})$ with $\Delta$ the frequency detuning between the emitter and the band center. In the resonant case $\Delta=0$, with the matching condition $g_{N'}/g_{N}=\beta^{-D}$, the self-energy simplifies to $\Sigma_{b}(\Delta=0)\approx\pm ig_{N}^{2}\mleft[1+\beta^{-2D}+(\pm i)^{D}\mleft(1+\beta^{-2D}\mright)\mright]/(2\sqrt{t_{R}t_{L}})$, which correctly predicts the dynamics in \figpanels{Fbasic}{b}{c}.

\emph{Nonreciprocal DFI.}---One of the most important applications of CGIEs is in-band DFI between giant emitters~\cite{braided,NoriGA,FCdeco,AFKchiral,complexDFI}. To explore whether this unique phenomenon is possible in the non-Bloch case, we extend the model in Eq.~(\ref{basicH}) to include an additional bosonic mode $c$, with the coupling points of the two modes arranged in a braided structure. In the interaction picture, the Hamiltonian of this extended model is
\begin{equation}
\begin{split}
H'&=\sum_{n=0}^{M-1} \mleft( t_{R}a_{n+1}^{\dag}a_{n} + t_{L}a_{n}^{\dag}a_{n+1} \mright) + \mleft( g_{N}b^{\dag}a_{N}\mright.\\
&\mleft.\quad\,+ g_{N+2}b^{\dag}a_{N+2} + g_{N+1}c^{\dag}a_{N+1} \mright.\\
&\mleft.\quad\,+ g_{N+3}c^{\dag}a_{N+3} + \text{H.c.} \mright),
\end{split}
\label{dimerH}
\end{equation}
where we assume that $b$ is coupled to sites $a_{N}$ and $a_{N+2}$ (with coupling strengths $g_{N}$ and $g_{N+2}$, respectively) and $c$ is coupled to sites $a_{N+1}$ and $a_{N+3}$ (with coupling strengths $g_{N+1}$ and $g_{N+3}$, respectively). Drawing on the result in Fig.~\ref{Fbasic}, we set $g_{N}=\beta^{2}g_{N+2}$ and $g_{N+1}=\beta^{2}g_{N+3}$ so that $b$ and $c$ almost do not decay into the lattice. 

\begin{figure}
\centering
\includegraphics[width=\linewidth]{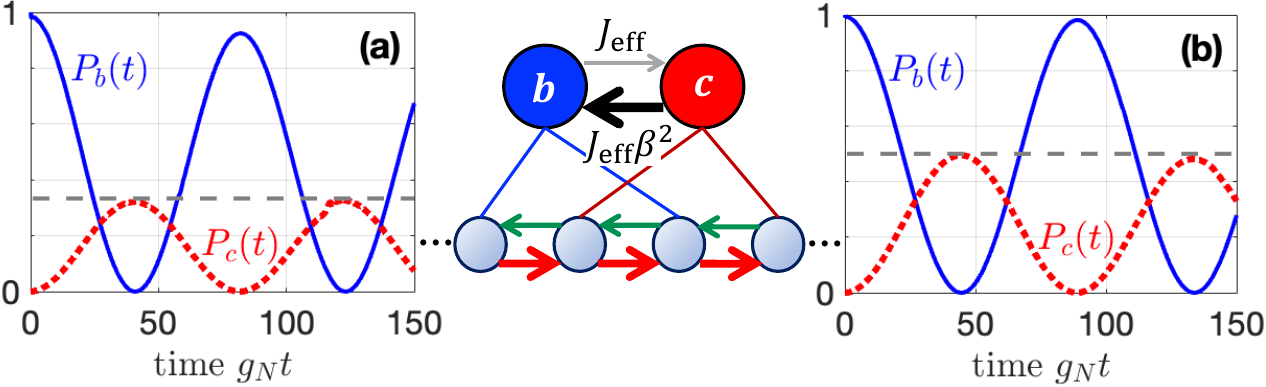}
\caption{Time evolution of $P_{b}(t)$ and $P_{c}(t)$ with (a) $\nu=20$, $\gamma=10$, and $\beta=\sqrt{3}$; (b) $\nu=15$, $\gamma=5$ and $\beta=\sqrt{2}$. The gray dashed lines correspond to the reference values $1/3$ in (a) and $1/2$ in (b). The middle inset is a schematic of the nonreciprocal DFI between $b$ and $c$ and the braided coupling structure. Other parameters are $g_{N}=g_{N+1}=1$, $g_{N+2}=g_{N+3}=\beta^{-2}$, and $M=1000$.}
\label{Fdimer}
\end{figure}

Figures~\figpanelNoPrefix{Fdimer}{a} and \figpanelNoPrefix{Fdimer}{b} depict the evolution of $P_{b}(t)=|u_{b}(t)|^{2}$ and $P_{c}(t)=|u_{c}(t)|^{2}$, with system state $|\psi_{2}(t)\rangle=[u_{b}(t)b^{\dag}+u_{c}(t)c^{\dag}+\sum_{n}u_{a,n}(t)a_{n}^{\dag}]|\emptyset\rangle$, initial condition $u_{\alpha}(0)=\delta_{\alpha,b}$, and different values of $\beta$. We find that $b$ and $c$ exchange energy in a nearly decoherence-free manner (there is a very weak decay due to the finite retardation effect) if the coupling strengths are matched as discussed above. In sharp contrast to the Hermitian case, the interaction here is nonreciprocal: the excitation is attenuated (amplified) when traveling from $b$ to $c$ (from $c$ to $b$). The attenuation/amplification is given by $\beta^{-2}$, as shown by the gray dashed lines in Fig.~\ref{Fdimer}, which is determined by the non-Bloch phase factor as well as the coupling separation of each giant emitter.

This phenomenon can be understood from the interaction parts of the self-energy of the emitters, which are obtained (see Sec.~II of~\cite{SM} for details) as
\begin{equation}
\Sigma_{bc}(0+i0^{+})\approx\frac{-G_{N}^{2}}{t_{R}}, \quad \Sigma_{cb}(0+i0^{+})\approx\frac{-G_{N}^{2}}{\beta^{2}t_{R}}.\label{selfbc}
\end{equation}
Since the real and imaginary parts of $\Sigma_{bc(cb)}(z)$ represent the effective coherent interaction from $c$ to $b$ ($b$ to $c$) and the collective relaxation of the two emitters, respectively, it is clear that $b$ and $c$ interact with each other in a decoherence-free but nonreciprocal manner, with a strength ratio $\beta^{-2}$ for the two directions. Most interestingly, the nonreciprocity of the interaction is \emph{opposite} to that of the bath: while the HN model amplifies right-moving fields, the emitter pair has a stronger interaction towards the left ($c$ is placed to the right of $b$). As we show in Sec.~III of~\cite{SM}, this reversed nonreciprocity can be understood in an intuitive picture based on hidden bound states~\cite{GongHN,GongHN2} induced by the interaction between the emitters and the HN model.


\emph{Absolutely unstable regime}---So far, we have focused on the ``convectively unstable regime'' with $t_{R}>t_{L}>0$~\cite{LonghiQD}, where the HN model can be mapped to a pseudo-Hermitian lattice subject to an imaginary gauge field (see Sec.~I of~\cite{SM}). In fact, there is a transition point $t_{L}=0$ (in this work, we always assume $t_{R}>0$) beyond which the HN model enters the ``absolutely unstable regime'' (where $t_{L}<0$). In the convectively unstable regime, a small emitter displays a complete (fractional) decay if its frequency is inside (outside) the energy band of the lattice. In the absolutely unstable regime, however, a small emitter has a secular pseudo-exponential energy growth~\cite{LonghiQD}. This phenomenon can be interpreted as a competition between the amplification and the effective transport of the field in the HN model (see Sec.~IV of~\cite{SM}): in the absolutely unstable regime, there is always a residual excitation component at the coupling point, suffering rapid amplification before being transferred away.

\begin{figure}
\centering
\includegraphics[width=\linewidth]{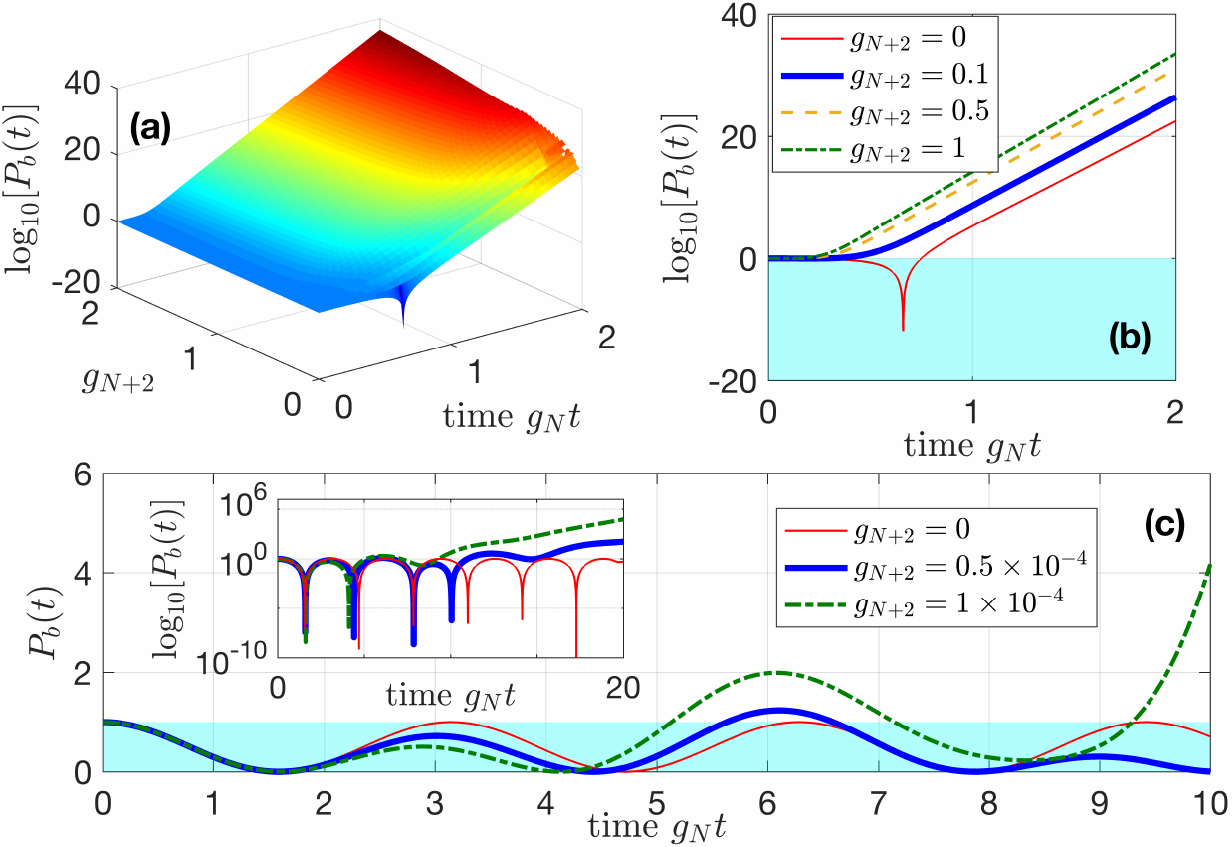}
\caption{Time evolution of $P_{b}(t)$ in (a, b) the absolutely unstable regime and (c) at the transition point $t_{L}=0$, for different values of $g_{N+2}$. In (a, b), $\gamma=20.5$; in (c), $\gamma=20$. The cyan rectangles in (b) and (c) indicate the areas with no energy growth. Other parameters are $g_{N}=1$, $\nu=20$, and $M=1000$.}
\label{Fabs}
\end{figure}

In contrast to the small-emitter case, a giant emitter can exhibit a secular growth even in the convectively unstable regime [see \figpanel{Fbasic}{b}]. This is due to the reabsorption of excitations emitted from another coupling point at an earlier moment and amplified by traveling in the HN model. This amplification mechanism, which arises from the combination of the giant-atom interference effect and the non-Hermiticity, is essentially different from that of a small emitter in the absolutely unstable regime.

We find that the CGIEs in Fig.~\ref{Fbasic} \emph{cannot} be realized in the absolutely unstable regime. Instead, a giant emitter behaves somewhat like a small one in the sense of the secular amplification. As shown in Fig.~\ref{Fabs}, the giant emitter always shows a pseudo-exponential energy growth; the growth rate increases monotonically with the coupling strength $g_{N+2}$ ($D=2$ is fixed). This is a bit different from the small-emitter case, where the emitter decays considerably before the secular growth.

Giant and small emitters also exhibit very different behaviors at the transition point $t_{L}=0$. As shown in \figpanel{Fabs}{c}, a small emitter simply exchanges energy with the coupled site since their dynamics is decoupled from the rest of the system~\cite{LonghiQD}, whereas a giant emitter shows energy growth since $\beta$ diverges at this point and the coupling matching condition $g_{N}/g_{N'}=\beta^{D}$ thus cannot be fulfilled.

\begin{figure}
\centering
\includegraphics[width=\linewidth]{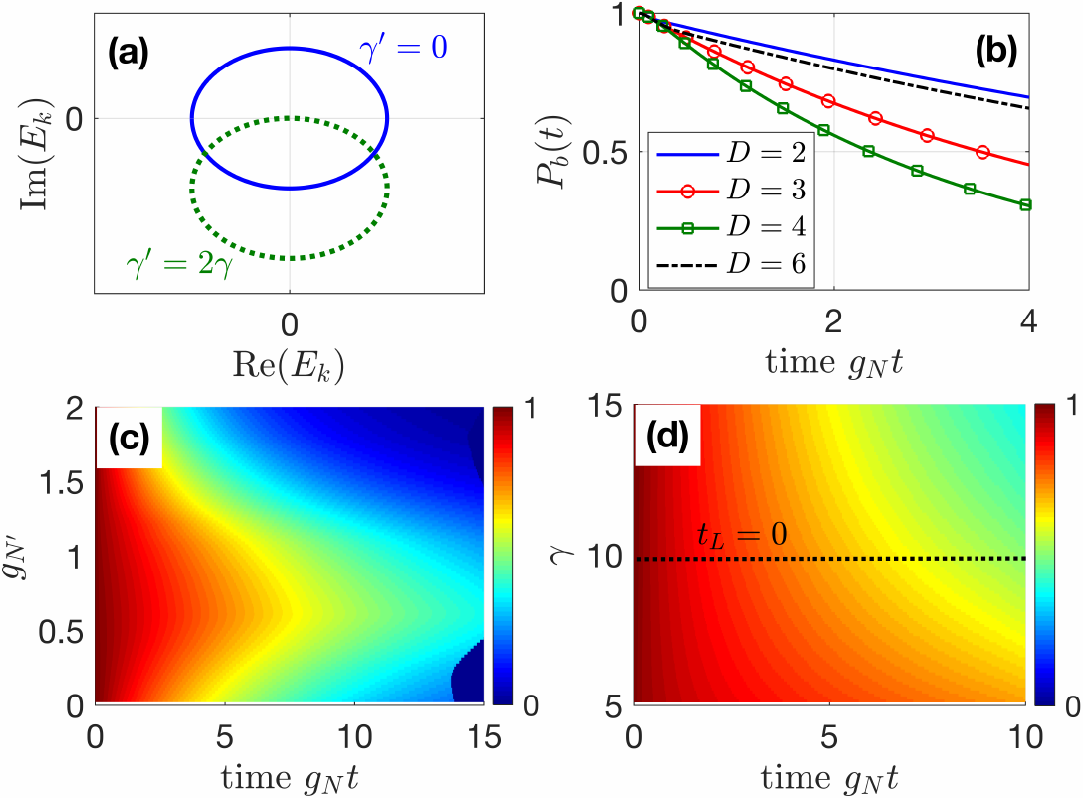}
\caption{(a) Spectra of HN models in the complex plane with and without on-site losses. (b)-(d) Time evolution of $P_{b}(t)$ in the stable regime for different values of (b) coupling separation $D$, (c) coupling strength $g_{N'}$, and (d) non-Hermiticity $\gamma$. We assume $g_{N'}=g_{N}$ and $\gamma=5$ in (b), $D=2$ and $\gamma=5$ in (c), and $g_{N'}=g_{N}$ and $D=2$ in (d). Other parameters are $g_{N}=1$, $\nu=10$, and $M=1000$.}
\label{Fstable}
\end{figure}

\emph{Giant emitter coupled to a stable HN model.}---A standard HN model is unstable (convectively or absolutely). Its spectrum forms a loop in the complex plane, with the imaginary part positive for some values of $k$ [see the blue solid loop in \figpanel{Fstable}{a}]. However, here we demonstrate how a giant emitter behaves if it is coupled to a stable HN model with large enough on-site losses~\cite{GongHN,GongHN2}.

To this end, we assume $H\rightarrow H+H_{\text{loss}}$ in Eq.~(\ref{basicH}) with $H_{\text{loss}}=-2i\sum_{n}\gamma a_{n}^{\dag}a_{n}$ describing uniform on-site losses of the lattice. As shown in \figpanel{Fstable}{a}, the spectrum of the HN model is within the lower half-plane in this case (green dotted loop). This implies that the whole system becomes stable, without field amplification in the lattice.


Figure~\figpanelNoPrefix{Fstable}{b} shows the time evolution of $P_{b}(t)$ for different values of $D=N'-N$. Similar to the Hermitian case~\cite{LonghiGA,ZhaoWbound,AFKstructured}, the relaxation dynamics of the emitter depend on the coupling separation $D$ (i.e., on the phase accumulation of emitted photons traveling between coupling points) with no need to match the nonlocal couplings. However, fractional decay of the giant emitter is unavailable in this case, regardless of the value of $g_{N'}$, as shown in \figpanel{Fstable}{c}. This can be understood from the bath now becoming purely dissipative so that the field always is attenuated when traveling between the coupling points. Thus, in a stable HN model, the giant emitter behaves as in a Bloch structured bath with uniform on-site losses. Moreover, \figpanel{Fstable}{d} shows that, in the stable regime, the giant emitter no longer exhibits qualitatively different dynamics for different signs of $t_{L}$, i.e., the transition point of the HN model disappears in this regime.

\emph{Conclusion and outlook.}---We have unveiled the self-interference effects of giant emitters coupled to an HN model, i.e., a structured bath featuring a non-Hermitian skin effect. The behaviors of giant emitters in this setting depend on the stability of the bath, which is closely related to the tunneling asymmetry. In the convectively unstable regime, where the HN model is equivalent to a pseudo-Hermitian lattice, CGIEs can be recovered by matching the nonlocal emitter-bath coupling strengths. In particular, we identify a non-Hermitian DFI condition under which the giant emitters show protected but nonreciprocal interactions. If the HN model enters the absolutely unstable regime, giant emitters exhibit secular energy growth just like the small emitters. An HN model made stable by introducing uniform on-site losses simply behaves like a Bloch (but dissipative) structured bath in the sense that giant emitters always show CGIEs with no need to match the nonlocal couplings.

Our findings can inspire a number of further investigations. For example, it has been shown recently that the HN model can be mapped to a two-dimensional hyperbolic lattice with the effective curvature determined by the non-Hermiticity $\gamma$~\cite{curveLv,DualityPRB}. It thus provides an exciting opportunity to explore giant-emitter physics in curved spaces without judiciously distorting the spatial configuration of a kagome lattice~\cite{Ccurve1,Ccurve2}. In Sec.~V of~\cite{SM}, we provide a brief discussion of designing a giant emitter coupled to such an effective hyperbolic lattice. Moreover, it is known that the spectra of some non-Hermitian topological insulators are sensitive to the boundary condition such that the conventional bulk-boundary correspondence no longer can predict the existence of their edge states~\cite{GongPRX,EpTopo,Yao2018,Yao20182,TopoOrigin,YFChen2020}. Studying how giant emitters affect these systems can be interesting since it shows the possibility of probing the topological phase and creating unconventional, topologically protected bound states. Considering the exotic behaviors of giant emitters in (Hermitian) two-dimensional lattices~\cite{AGT2D1,AGT2D2,AFK2D}, it could also be interesting to introduce the non-Hermitian skin effect to higher-dimensional baths~\cite{HigherOrderSkin,LuSkin2D} and study their interplay with giant emitters. The results, such as the nonreciprocal DFIs, hold promise for non-Hermitian many-body simulations due to the interaction protection mechanism in a non-Hermitian bath~\cite{FengNP2017}.

\acknowledgments{\emph{Acknowledgments.}---LD thanks Francesco Ciccarello and Xin Wang for helpful discussions. YZ is supported by the National Natural Science Foundation of China (under Grant No. 12074061). AFK acknowledges support from the Swedish Research Council (grant number 2019-03696), from the Swedish Foundation for Strategic Research, and from the Knut and Alice Wallenberg Foundation through the Wallenberg Centre for Quantum Technology (WACQT).}

\bibliography{GAskin_ref}

\begin{thebibliography}{81}%
\makeatletter
\providecommand \@ifxundefined [1]{%
 \@ifx{#1\undefined}
}%
\providecommand \@ifnum [1]{%
 \ifnum #1\expandafter \@firstoftwo
 \else \expandafter \@secondoftwo
 \fi
}%
\providecommand \@ifx [1]{%
 \ifx #1\expandafter \@firstoftwo
 \else \expandafter \@secondoftwo
 \fi
}%
\providecommand \natexlab [1]{#1}%
\providecommand \enquote  [1]{``#1''}%
\providecommand \bibnamefont  [1]{#1}%
\providecommand \bibfnamefont [1]{#1}%
\providecommand \citenamefont [1]{#1}%
\providecommand \href@noop [0]{\@secondoftwo}%
\providecommand \href [0]{\begingroup \@sanitize@url \@href}%
\providecommand \@href[1]{\@@startlink{#1}\@@href}%
\providecommand \@@href[1]{\endgroup#1\@@endlink}%
\providecommand \@sanitize@url [0]{\catcode `\\12\catcode `\$12\catcode
  `\&12\catcode `\#12\catcode `\^12\catcode `\_12\catcode `\%12\relax}%
\providecommand \@@startlink[1]{}%
\providecommand \@@endlink[0]{}%
\providecommand \url  [0]{\begingroup\@sanitize@url \@url }%
\providecommand \@url [1]{\endgroup\@href {#1}{\urlprefix }}%
\providecommand \urlprefix  [0]{URL }%
\providecommand \Eprint [0]{\href }%
\providecommand \doibase [0]{https://doi.org/}%
\providecommand \selectlanguage [0]{\@gobble}%
\providecommand \bibinfo  [0]{\@secondoftwo}%
\providecommand \bibfield  [0]{\@secondoftwo}%
\providecommand \translation [1]{[#1]}%
\providecommand \BibitemOpen [0]{}%
\providecommand \bibitemStop [0]{}%
\providecommand \bibitemNoStop [0]{.\EOS\space}%
\providecommand \EOS [0]{\spacefactor3000\relax}%
\providecommand \BibitemShut  [1]{\csname bibitem#1\endcsname}%
\let\auto@bib@innerbib\@empty
\bibitem [{\citenamefont {Kockum}(2021)}]{fiveyear}%
  \BibitemOpen
  \bibfield  {author} {\bibinfo {author} {\bibfnamefont {A.~F.}\ \bibnamefont
  {Kockum}},\ }\bibfield  {title} {\bibinfo {title} {Quantum {O}ptics with
  {G}iant {A}toms---the {F}irst {F}ive {Y}ears},\ }in\ \href@noop {} {\emph
  {\bibinfo {booktitle} {International Symposium on Mathematics, Quantum
  Theory, and Cryptography}}},\ \bibinfo {editor} {edited by\ \bibinfo {editor}
  {\bibfnamefont {T.}~\bibnamefont {Takagi}}, \bibinfo {editor} {\bibfnamefont
  {M.}~\bibnamefont {Wakayama}}, \bibinfo {editor} {\bibfnamefont
  {K.}~\bibnamefont {Tanaka}}, \bibinfo {editor} {\bibfnamefont
  {N.}~\bibnamefont {Kunihiro}}, \bibinfo {editor} {\bibfnamefont
  {K.}~\bibnamefont {Kimoto}},\ and\ \bibinfo {editor} {\bibfnamefont
  {Y.}~\bibnamefont {Ikematsu}}}\ (\bibinfo  {publisher} {Springer Singapore},\
  \bibinfo {address} {Singapore},\ \bibinfo {year} {2021})\ pp.\ \bibinfo
  {pages} {125--146}\BibitemShut {NoStop}%
\bibitem [{\citenamefont {Kockum}\ \emph {et~al.}(2014)\citenamefont {Kockum},
  \citenamefont {Delsing},\ and\ \citenamefont {Johansson}}]{LambAFK}%
  \BibitemOpen
  \bibfield  {author} {\bibinfo {author} {\bibfnamefont {A.~F.}\ \bibnamefont
  {Kockum}}, \bibinfo {author} {\bibfnamefont {P.}~\bibnamefont {Delsing}},\
  and\ \bibinfo {author} {\bibfnamefont {G.}~\bibnamefont {Johansson}},\
  }\bibfield  {title} {\bibinfo {title} {Designing frequency-dependent
  relaxation rates and {L}amb shifts for a giant artificial atom},\ }\href
  {https://journals.aps.org/pra/abstract/10.1103/PhysRevA.90.013837} {\bibfield
   {journal} {\bibinfo  {journal} {Phys. Rev. A}\ }\textbf {\bibinfo {volume}
  {90}},\ \bibinfo {pages} {013837} (\bibinfo {year} {2014})}\BibitemShut
  {NoStop}%
\bibitem [{\citenamefont {Gustafsson}\ \emph {et~al.}(2014)\citenamefont
  {Gustafsson}, \citenamefont {Aref}, \citenamefont {Kockum}, \citenamefont
  {Ekstr\"{o}m}, \citenamefont {Johansson},\ and\ \citenamefont
  {Delsing}}]{SAW2014}%
  \BibitemOpen
  \bibfield  {author} {\bibinfo {author} {\bibfnamefont {M.~V.}\ \bibnamefont
  {Gustafsson}}, \bibinfo {author} {\bibfnamefont {T.}~\bibnamefont {Aref}},
  \bibinfo {author} {\bibfnamefont {A.~F.}\ \bibnamefont {Kockum}}, \bibinfo
  {author} {\bibfnamefont {M.~K.}\ \bibnamefont {Ekstr\"{o}m}}, \bibinfo
  {author} {\bibfnamefont {G.}~\bibnamefont {Johansson}},\ and\ \bibinfo
  {author} {\bibfnamefont {P.}~\bibnamefont {Delsing}},\ }\bibfield  {title}
  {\bibinfo {title} {Propagating phonons coupled to an artificial atom},\
  }\href {https://www.science.org/doi/10.1126/science.1257219} {\bibfield
  {journal} {\bibinfo  {journal} {Science}\ }\textbf {\bibinfo {volume}
  {346}},\ \bibinfo {pages} {207} (\bibinfo {year} {2014})}\BibitemShut
  {NoStop}%
\bibitem [{\citenamefont {Guo}\ \emph {et~al.}(2017)\citenamefont {Guo},
  \citenamefont {Grimsmo}, \citenamefont {Kockum}, \citenamefont {Pletyukhov},\
  and\ \citenamefont {Johansson}}]{GLZ2017}%
  \BibitemOpen
  \bibfield  {author} {\bibinfo {author} {\bibfnamefont {L.}~\bibnamefont
  {Guo}}, \bibinfo {author} {\bibfnamefont {A.}~\bibnamefont {Grimsmo}},
  \bibinfo {author} {\bibfnamefont {A.~F.}\ \bibnamefont {Kockum}}, \bibinfo
  {author} {\bibfnamefont {M.}~\bibnamefont {Pletyukhov}},\ and\ \bibinfo
  {author} {\bibfnamefont {G.}~\bibnamefont {Johansson}},\ }\bibfield  {title}
  {\bibinfo {title} {Giant acoustic atom: A single quantum system with a
  deterministic time delay},\ }\href
  {https://journals.aps.org/pra/abstract/10.1103/PhysRevA.95.053821} {\bibfield
   {journal} {\bibinfo  {journal} {Phys. Rev. A}\ }\textbf {\bibinfo {volume}
  {95}},\ \bibinfo {pages} {053821} (\bibinfo {year} {2017})}\BibitemShut
  {NoStop}%
\bibitem [{\citenamefont {Andersson}\ \emph {et~al.}(2019)\citenamefont
  {Andersson}, \citenamefont {Suri}, \citenamefont {Guo}, \citenamefont
  {Aref},\ and\ \citenamefont {Delsing}}]{nonexp}%
  \BibitemOpen
  \bibfield  {author} {\bibinfo {author} {\bibfnamefont {G.}~\bibnamefont
  {Andersson}}, \bibinfo {author} {\bibfnamefont {B.}~\bibnamefont {Suri}},
  \bibinfo {author} {\bibfnamefont {L.}~\bibnamefont {Guo}}, \bibinfo {author}
  {\bibfnamefont {T.}~\bibnamefont {Aref}},\ and\ \bibinfo {author}
  {\bibfnamefont {P.}~\bibnamefont {Delsing}},\ }\bibfield  {title} {\bibinfo
  {title} {Non-exponential decay of a giant artificial atom},\ }\href
  {https://www.nature.com/articles/s41567-019-0605-6} {\bibfield  {journal}
  {\bibinfo  {journal} {Nat. Phys.}\ }\textbf {\bibinfo {volume} {15}},\
  \bibinfo {pages} {1123} (\bibinfo {year} {2019})}\BibitemShut {NoStop}%
\bibitem [{\citenamefont {Vadiraj}\ \emph {et~al.}(2021)\citenamefont
  {Vadiraj}, \citenamefont {Ask}, \citenamefont {McConkey}, \citenamefont
  {Nsanzineza}, \citenamefont {Sandbo~Chang}, \citenamefont {Kockum},\ and\
  \citenamefont {Wilson}}]{WilsonPRA2021}%
  \BibitemOpen
  \bibfield  {author} {\bibinfo {author} {\bibfnamefont {A.~M.}\ \bibnamefont
  {Vadiraj}}, \bibinfo {author} {\bibfnamefont {A.}~\bibnamefont {Ask}},
  \bibinfo {author} {\bibfnamefont {T.~G.}\ \bibnamefont {McConkey}}, \bibinfo
  {author} {\bibfnamefont {I.}~\bibnamefont {Nsanzineza}}, \bibinfo {author}
  {\bibfnamefont {C.~W.}\ \bibnamefont {Sandbo~Chang}}, \bibinfo {author}
  {\bibfnamefont {A.~F.}\ \bibnamefont {Kockum}},\ and\ \bibinfo {author}
  {\bibfnamefont {C.~M.}\ \bibnamefont {Wilson}},\ }\bibfield  {title}
  {\bibinfo {title} {Engineering the level structure of a giant artificial atom
  in waveguide quantum electrodynamics},\ }\href
  {https://journals.aps.org/pra/abstract/10.1103/PhysRevA.103.023710}
  {\bibfield  {journal} {\bibinfo  {journal} {Phys. Rev. A}\ }\textbf {\bibinfo
  {volume} {103}},\ \bibinfo {pages} {023710} (\bibinfo {year}
  {2021})}\BibitemShut {NoStop}%
\bibitem [{\citenamefont {Kannan}\ \emph {et~al.}(2020)\citenamefont {Kannan},
  \citenamefont {Ruckriegel}, \citenamefont {Campbell}, \citenamefont {Kockum},
  \citenamefont {Braum\"{u}ller}, \citenamefont {Kim}, \citenamefont
  {Kjaergaard}, \citenamefont {Krantz}, \citenamefont {Melville}, \citenamefont
  {Niedzielski}, \citenamefont {Veps\"{a}l\"{a}inen}, \citenamefont {Winik},
  \citenamefont {Yoder}, \citenamefont {Nori}, \citenamefont {Orlando},
  \citenamefont {Gustavsson},\ and\ \citenamefont {Oliver}}]{braided}%
  \BibitemOpen
  \bibfield  {author} {\bibinfo {author} {\bibfnamefont {B.}~\bibnamefont
  {Kannan}}, \bibinfo {author} {\bibfnamefont {M.}~\bibnamefont {Ruckriegel}},
  \bibinfo {author} {\bibfnamefont {D.}~\bibnamefont {Campbell}}, \bibinfo
  {author} {\bibfnamefont {A.~F.}\ \bibnamefont {Kockum}}, \bibinfo {author}
  {\bibfnamefont {J.}~\bibnamefont {Braum\"{u}ller}}, \bibinfo {author}
  {\bibfnamefont {D.}~\bibnamefont {Kim}}, \bibinfo {author} {\bibfnamefont
  {M.}~\bibnamefont {Kjaergaard}}, \bibinfo {author} {\bibfnamefont
  {P.}~\bibnamefont {Krantz}}, \bibinfo {author} {\bibfnamefont
  {A.}~\bibnamefont {Melville}}, \bibinfo {author} {\bibfnamefont {B.~M.}\
  \bibnamefont {Niedzielski}}, \bibinfo {author} {\bibfnamefont
  {A.}~\bibnamefont {Veps\"{a}l\"{a}inen}}, \bibinfo {author} {\bibfnamefont
  {R.}~\bibnamefont {Winik}}, \bibinfo {author} {\bibfnamefont
  {J.}~\bibnamefont {Yoder}}, \bibinfo {author} {\bibfnamefont
  {F.}~\bibnamefont {Nori}}, \bibinfo {author} {\bibfnamefont {T.~P.}\
  \bibnamefont {Orlando}}, \bibinfo {author} {\bibfnamefont {S.}~\bibnamefont
  {Gustavsson}},\ and\ \bibinfo {author} {\bibfnamefont {W.~D.}\ \bibnamefont
  {Oliver}},\ }\bibfield  {title} {\bibinfo {title} {Waveguide quantum
  electrodynamics with superconducting artificial giant atoms},\ }\href
  {https://www.nature.com/articles/s41586-020-2529-9} {\bibfield  {journal}
  {\bibinfo  {journal} {Nature (London)}\ }\textbf {\bibinfo {volume} {583}},\
  \bibinfo {pages} {775} (\bibinfo {year} {2020})}\BibitemShut {NoStop}%
\bibitem [{\citenamefont {Kockum}\ \emph {et~al.}(2018)\citenamefont {Kockum},
  \citenamefont {Johansson},\ and\ \citenamefont {Nori}}]{NoriGA}%
  \BibitemOpen
  \bibfield  {author} {\bibinfo {author} {\bibfnamefont {A.~F.}\ \bibnamefont
  {Kockum}}, \bibinfo {author} {\bibfnamefont {G.}~\bibnamefont {Johansson}},\
  and\ \bibinfo {author} {\bibfnamefont {F.}~\bibnamefont {Nori}},\ }\bibfield
  {title} {\bibinfo {title} {Decoherence-{F}ree {I}nteraction between {G}iant
  {A}toms in {W}aveguide {Q}uantum {E}lectrodynamics},\ }\href
  {https://journals.aps.org/prl/abstract/10.1103/PhysRevLett.120.140404}
  {\bibfield  {journal} {\bibinfo  {journal} {Phys. Rev. Lett.}\ }\textbf
  {\bibinfo {volume} {120}},\ \bibinfo {pages} {140404} (\bibinfo {year}
  {2018})}\BibitemShut {NoStop}%
\bibitem [{\citenamefont {Carollo}\ \emph {et~al.}(2020)\citenamefont
  {Carollo}, \citenamefont {Cilluffo},\ and\ \citenamefont
  {Ciccarello}}]{FCdeco}%
  \BibitemOpen
  \bibfield  {author} {\bibinfo {author} {\bibfnamefont {A.}~\bibnamefont
  {Carollo}}, \bibinfo {author} {\bibfnamefont {D.}~\bibnamefont {Cilluffo}},\
  and\ \bibinfo {author} {\bibfnamefont {F.}~\bibnamefont {Ciccarello}},\
  }\bibfield  {title} {\bibinfo {title} {Mechanism of decoherence-free coupling
  between giant atoms},\ }\href
  {https://journals.aps.org/prresearch/abstract/10.1103/PhysRevResearch.2.043184}
  {\bibfield  {journal} {\bibinfo  {journal} {Phys. Rev. Res.}\ }\textbf
  {\bibinfo {volume} {2}},\ \bibinfo {pages} {043184} (\bibinfo {year}
  {2020})}\BibitemShut {NoStop}%
\bibitem [{\citenamefont {Soro}\ and\ \citenamefont
  {Kockum}(2022)}]{AFKchiral}%
  \BibitemOpen
  \bibfield  {author} {\bibinfo {author} {\bibfnamefont {A.}~\bibnamefont
  {Soro}}\ and\ \bibinfo {author} {\bibfnamefont {A.~F.}\ \bibnamefont
  {Kockum}},\ }\bibfield  {title} {\bibinfo {title} {Chiral quantum optics with
  giant atoms},\ }\href
  {https://journals.aps.org/pra/abstract/10.1103/PhysRevA.105.023712}
  {\bibfield  {journal} {\bibinfo  {journal} {Phys. Rev. A}\ }\textbf {\bibinfo
  {volume} {105}},\ \bibinfo {pages} {023712} (\bibinfo {year}
  {2022})}\BibitemShut {NoStop}%
\bibitem [{\citenamefont {Du}\ \emph {et~al.}(2023{\natexlab{a}})\citenamefont
  {Du}, \citenamefont {Guo},\ and\ \citenamefont {Li}}]{complexDFI}%
  \BibitemOpen
  \bibfield  {author} {\bibinfo {author} {\bibfnamefont {L.}~\bibnamefont
  {Du}}, \bibinfo {author} {\bibfnamefont {L.}~\bibnamefont {Guo}},\ and\
  \bibinfo {author} {\bibfnamefont {Y.}~\bibnamefont {Li}},\ }\bibfield
  {title} {\bibinfo {title} {Complex decoherence-free interactions between
  giant atoms},\ }\href
  {https://journals.aps.org/pra/abstract/10.1103/PhysRevA.107.023705}
  {\bibfield  {journal} {\bibinfo  {journal} {Phys. Rev. A}\ }\textbf {\bibinfo
  {volume} {107}},\ \bibinfo {pages} {023705} (\bibinfo {year}
  {2023}{\natexlab{a}})}\BibitemShut {NoStop}%
\bibitem [{\citenamefont {Bay}\ \emph {et~al.}(1997)\citenamefont {Bay},
  \citenamefont {Lambropoulos},\ and\ \citenamefont {M\"{o}lmer}}]{SDFI1}%
  \BibitemOpen
  \bibfield  {author} {\bibinfo {author} {\bibfnamefont {S.}~\bibnamefont
  {Bay}}, \bibinfo {author} {\bibfnamefont {P.}~\bibnamefont {Lambropoulos}},\
  and\ \bibinfo {author} {\bibfnamefont {K.}~\bibnamefont {M\"{o}lmer}},\
  }\bibfield  {title} {\bibinfo {title} {Atom-atom interaction in strongly
  modified reservoirs},\ }\href
  {https://journals.aps.org/pra/abstract/10.1103/PhysRevA.55.1485} {\bibfield
  {journal} {\bibinfo  {journal} {Phys. Rev. A}\ }\textbf {\bibinfo {volume}
  {55}},\ \bibinfo {pages} {1485} (\bibinfo {year} {1997})}\BibitemShut
  {NoStop}%
\bibitem [{\citenamefont {Lambropoulos}\ \emph {et~al.}(2000)\citenamefont
  {Lambropoulos}, \citenamefont {Nikolopoulos}, \citenamefont {Nielsen},\ and\
  \citenamefont {Bay}}]{SDFI2}%
  \BibitemOpen
  \bibfield  {author} {\bibinfo {author} {\bibfnamefont {P.}~\bibnamefont
  {Lambropoulos}}, \bibinfo {author} {\bibfnamefont {G.~M.}\ \bibnamefont
  {Nikolopoulos}}, \bibinfo {author} {\bibfnamefont {T.~R.}\ \bibnamefont
  {Nielsen}},\ and\ \bibinfo {author} {\bibfnamefont {S.}~\bibnamefont {Bay}},\
  }\bibfield  {title} {\bibinfo {title} {Fundamental quantum optics in
  structured reservoirs},\ }\href
  {https://iopscience.iop.org/article/10.1088/0034-4885/63/4/201} {\bibfield
  {journal} {\bibinfo  {journal} {Rep. Prog. Phys.}\ }\textbf {\bibinfo
  {volume} {63}},\ \bibinfo {pages} {455} (\bibinfo {year} {2000})}\BibitemShut
  {NoStop}%
\bibitem [{\citenamefont {Shahmoon}\ and\ \citenamefont
  {Kurizki}(2013)}]{SDFI3}%
  \BibitemOpen
  \bibfield  {author} {\bibinfo {author} {\bibfnamefont {E.}~\bibnamefont
  {Shahmoon}}\ and\ \bibinfo {author} {\bibfnamefont {G.}~\bibnamefont
  {Kurizki}},\ }\bibfield  {title} {\bibinfo {title} {Nonradiative interaction
  and entanglement between distant atoms},\ }\href
  {https://journals.aps.org/pra/abstract/10.1103/PhysRevA.87.033831} {\bibfield
   {journal} {\bibinfo  {journal} {Phys. Rev. A}\ }\textbf {\bibinfo {volume}
  {87}},\ \bibinfo {pages} {033831} (\bibinfo {year} {2013})}\BibitemShut
  {NoStop}%
\bibitem [{\citenamefont {Wang}\ \emph {et~al.}(2020)\citenamefont {Wang},
  \citenamefont {Liu}, \citenamefont {Kockum}, \citenamefont {Li},\ and\
  \citenamefont {Nori}}]{WXchiral1}%
  \BibitemOpen
  \bibfield  {author} {\bibinfo {author} {\bibfnamefont {X.}~\bibnamefont
  {Wang}}, \bibinfo {author} {\bibfnamefont {T.}~\bibnamefont {Liu}}, \bibinfo
  {author} {\bibfnamefont {A.~F.}\ \bibnamefont {Kockum}}, \bibinfo {author}
  {\bibfnamefont {H.-R.}\ \bibnamefont {Li}},\ and\ \bibinfo {author}
  {\bibfnamefont {F.}~\bibnamefont {Nori}},\ }\bibfield  {title} {\bibinfo
  {title} {Tunable {C}hiral {B}ound {S}tates with {G}iant {A}toms},\ }\href
  {https://journals.aps.org/prl/abstract/10.1103/PhysRevLett.126.043602}
  {\bibfield  {journal} {\bibinfo  {journal} {Phys. Rev. Lett.}\ }\textbf
  {\bibinfo {volume} {126}},\ \bibinfo {pages} {043602} (\bibinfo {year}
  {2020})}\BibitemShut {NoStop}%
\bibitem [{\citenamefont {Wang}\ and\ \citenamefont {r.~Li}(2022)}]{WXchiral2}%
  \BibitemOpen
  \bibfield  {author} {\bibinfo {author} {\bibfnamefont {X.}~\bibnamefont
  {Wang}}\ and\ \bibinfo {author} {\bibfnamefont {H.}~\bibnamefont {r.~Li}},\
  }\bibfield  {title} {\bibinfo {title} {Chiral quantum network with giant
  atoms},\ }\href
  {https://iopscience.iop.org/article/10.1088/2058-9565/ac6a04#qstac6a04app1}
  {\bibfield  {journal} {\bibinfo  {journal} {Quantum Sci. Technol.}\ }\textbf
  {\bibinfo {volume} {7}},\ \bibinfo {pages} {035007} (\bibinfo {year}
  {2022})}\BibitemShut {NoStop}%
\bibitem [{\citenamefont {Du}\ \emph {et~al.}(2022{\natexlab{a}})\citenamefont
  {Du}, \citenamefont {Zhang}, \citenamefont {Wu}, \citenamefont {Kockum},\
  and\ \citenamefont {Li}}]{DLprl}%
  \BibitemOpen
  \bibfield  {author} {\bibinfo {author} {\bibfnamefont {L.}~\bibnamefont
  {Du}}, \bibinfo {author} {\bibfnamefont {Y.}~\bibnamefont {Zhang}}, \bibinfo
  {author} {\bibfnamefont {J.-H.}\ \bibnamefont {Wu}}, \bibinfo {author}
  {\bibfnamefont {A.~F.}\ \bibnamefont {Kockum}},\ and\ \bibinfo {author}
  {\bibfnamefont {Y.}~\bibnamefont {Li}},\ }\bibfield  {title} {\bibinfo
  {title} {Giant {A}toms in a {S}ynthetic {F}requency {D}imension},\ }\href
  {https://journals.aps.org/prl/abstract/10.1103/PhysRevLett.128.223602}
  {\bibfield  {journal} {\bibinfo  {journal} {Phys. Rev. Lett.}\ }\textbf
  {\bibinfo {volume} {128}},\ \bibinfo {pages} {223602} (\bibinfo {year}
  {2022}{\natexlab{a}})}\BibitemShut {NoStop}%
\bibitem [{\citenamefont {Chen}\ \emph {et~al.}(2022)\citenamefont {Chen},
  \citenamefont {Du}, \citenamefont {Guo}, \citenamefont {Wang}, \citenamefont
  {Zhang}, \citenamefont {Li},\ and\ \citenamefont {Wu}}]{CYTcp}%
  \BibitemOpen
  \bibfield  {author} {\bibinfo {author} {\bibfnamefont {Y.-T.}\ \bibnamefont
  {Chen}}, \bibinfo {author} {\bibfnamefont {L.}~\bibnamefont {Du}}, \bibinfo
  {author} {\bibfnamefont {L.}~\bibnamefont {Guo}}, \bibinfo {author}
  {\bibfnamefont {Z.}~\bibnamefont {Wang}}, \bibinfo {author} {\bibfnamefont
  {Y.}~\bibnamefont {Zhang}}, \bibinfo {author} {\bibfnamefont
  {Y.}~\bibnamefont {Li}},\ and\ \bibinfo {author} {\bibfnamefont {J.-H.}\
  \bibnamefont {Wu}},\ }\bibfield  {title} {\bibinfo {title} {Nonreciprocal and
  chiral single-photon scattering for giant atoms},\ }\href
  {https://www.nature.com/articles/s42005-022-00991-3} {\bibfield  {journal}
  {\bibinfo  {journal} {Commun. Physc.}\ }\textbf {\bibinfo {volume} {5}},\
  \bibinfo {pages} {215} (\bibinfo {year} {2022})}\BibitemShut {NoStop}%
\bibitem [{\citenamefont {Du}\ \emph {et~al.}(2023{\natexlab{b}})\citenamefont
  {Du}, \citenamefont {Chen}, \citenamefont {Zhang}, \citenamefont {Li},\ and\
  \citenamefont {Wu}}]{LeiQST}%
  \BibitemOpen
  \bibfield  {author} {\bibinfo {author} {\bibfnamefont {L.}~\bibnamefont
  {Du}}, \bibinfo {author} {\bibfnamefont {Y.-T.}\ \bibnamefont {Chen}},
  \bibinfo {author} {\bibfnamefont {Y.}~\bibnamefont {Zhang}}, \bibinfo
  {author} {\bibfnamefont {Y.}~\bibnamefont {Li}},\ and\ \bibinfo {author}
  {\bibfnamefont {J.-H.}\ \bibnamefont {Wu}},\ }\bibfield  {title} {\bibinfo
  {title} {Decay dynamics of a giant atom in a structured bath with broken
  time-reversal symmetry},\ }\href
  {https://iopscience.iop.org/article/10.1088/2058-9565/ace54c} {\bibfield
  {journal} {\bibinfo  {journal} {Quantum Sci. Technol.}\ }\textbf {\bibinfo
  {volume} {8}},\ \bibinfo {pages} {045010} (\bibinfo {year}
  {2023}{\natexlab{b}})}\BibitemShut {NoStop}%
\bibitem [{\citenamefont {Joshi}\ \emph {et~al.}(2023)\citenamefont {Joshi},
  \citenamefont {Yang},\ and\ \citenamefont {Mirhosseini}}]{MohammadArxiv}%
  \BibitemOpen
  \bibfield  {author} {\bibinfo {author} {\bibfnamefont {C.}~\bibnamefont
  {Joshi}}, \bibinfo {author} {\bibfnamefont {F.}~\bibnamefont {Yang}},\ and\
  \bibinfo {author} {\bibfnamefont {M.}~\bibnamefont {Mirhosseini}},\
  }\bibfield  {title} {\bibinfo {title} {Resonance {F}luorescence of a {C}hiral
  {A}rtificial {A}tom},\ }\href
  {https://journals.aps.org/prx/abstract/10.1103/PhysRevX.13.021039} {\bibfield
   {journal} {\bibinfo  {journal} {Phys. Rev. X}\ }\textbf {\bibinfo {volume}
  {13}},\ \bibinfo {pages} {021039} (\bibinfo {year} {2023})}\BibitemShut
  {NoStop}%
\bibitem [{\citenamefont {Kittel}(2005)}]{Bloch}%
  \BibitemOpen
  \bibfield  {author} {\bibinfo {author} {\bibfnamefont {C.}~\bibnamefont
  {Kittel}},\ }\href@noop {} {\emph {\bibinfo {title} {Introduction to Solid
  State Physics}}}\ (\bibinfo  {publisher} {Weily, New York},\ \bibinfo {year}
  {2005})\BibitemShut {NoStop}%
\bibitem [{\citenamefont {Bender}\ and\ \citenamefont
  {Boettcher}(1998)}]{NHg1}%
  \BibitemOpen
  \bibfield  {author} {\bibinfo {author} {\bibfnamefont {C.~M.}\ \bibnamefont
  {Bender}}\ and\ \bibinfo {author} {\bibfnamefont {S.}~\bibnamefont
  {Boettcher}},\ }\bibfield  {title} {\bibinfo {title} {Real {S}pectra in
  {N}on-{H}ermitian {H}amiltonians {H}aving $\mathcal{PT}$ {S}ymmetry},\ }\href
  {https://journals.aps.org/prl/abstract/10.1103/PhysRevLett.80.5243}
  {\bibfield  {journal} {\bibinfo  {journal} {Phys. Rev. Lett.}\ }\textbf
  {\bibinfo {volume} {80}},\ \bibinfo {pages} {5243} (\bibinfo {year}
  {1998})}\BibitemShut {NoStop}%
\bibitem [{\citenamefont {Bender}(2007)}]{NHg2}%
  \BibitemOpen
  \bibfield  {author} {\bibinfo {author} {\bibfnamefont {C.~M.}\ \bibnamefont
  {Bender}},\ }\bibfield  {title} {\bibinfo {title} {Making sense of
  non-{H}ermitian {H}amiltonians},\ }\href
  {https://iopscience.iop.org/article/10.1088/0034-4885/70/6/R03} {\bibfield
  {journal} {\bibinfo  {journal} {Rep. Prog. Phys.}\ }\textbf {\bibinfo
  {volume} {70}},\ \bibinfo {pages} {947} (\bibinfo {year} {2007})}\BibitemShut
  {NoStop}%
\bibitem [{\citenamefont {Longhi}(2009)}]{NHg3}%
  \BibitemOpen
  \bibfield  {author} {\bibinfo {author} {\bibfnamefont {S.}~\bibnamefont
  {Longhi}},\ }\bibfield  {title} {\bibinfo {title} {Bloch {O}scillations in
  {C}omplex {C}rystals with $\mathcal{PT}$ {S}ymmetry},\ }\href
  {https://journals.aps.org/prl/abstract/10.1103/PhysRevLett.103.123601}
  {\bibfield  {journal} {\bibinfo  {journal} {Phys. Rev. Lett.}\ }\textbf
  {\bibinfo {volume} {103}},\ \bibinfo {pages} {123601} (\bibinfo {year}
  {2009})}\BibitemShut {NoStop}%
\bibitem [{\citenamefont {Wu}\ \emph {et~al.}(2014)\citenamefont {Wu},
  \citenamefont {Artoni},\ and\ \citenamefont {La~Rocca}}]{NHg4}%
  \BibitemOpen
  \bibfield  {author} {\bibinfo {author} {\bibfnamefont {J.-H.}\ \bibnamefont
  {Wu}}, \bibinfo {author} {\bibfnamefont {M.}~\bibnamefont {Artoni}},\ and\
  \bibinfo {author} {\bibfnamefont {G.~C.}\ \bibnamefont {La~Rocca}},\
  }\bibfield  {title} {\bibinfo {title} {Non-{H}ermitian {D}egeneracies and
  {U}nidirectional {R}eflectionless {A}tomic {L}attices},\ }\href
  {https://journals.aps.org/prl/abstract/10.1103/PhysRevLett.113.123004}
  {\bibfield  {journal} {\bibinfo  {journal} {Phys. Rev. Lett.}\ }\textbf
  {\bibinfo {volume} {113}},\ \bibinfo {pages} {123004} (\bibinfo {year}
  {2014})}\BibitemShut {NoStop}%
\bibitem [{\citenamefont {Konotop}\ \emph {et~al.}(2016)\citenamefont
  {Konotop}, \citenamefont {Yang},\ and\ \citenamefont {Zezyulin}}]{NHg5}%
  \BibitemOpen
  \bibfield  {author} {\bibinfo {author} {\bibfnamefont {V.~V.}\ \bibnamefont
  {Konotop}}, \bibinfo {author} {\bibfnamefont {J.}~\bibnamefont {Yang}},\ and\
  \bibinfo {author} {\bibfnamefont {D.~A.}\ \bibnamefont {Zezyulin}},\
  }\bibfield  {title} {\bibinfo {title} {Nonlinear waves in
  $\mathcal{PT}$-symmetric systems},\ }\href
  {https://journals.aps.org/rmp/abstract/10.1103/RevModPhys.88.035002}
  {\bibfield  {journal} {\bibinfo  {journal} {Rev. Mod. Phys.}\ }\textbf
  {\bibinfo {volume} {88}},\ \bibinfo {pages} {035002} (\bibinfo {year}
  {2016})}\BibitemShut {NoStop}%
\bibitem [{\citenamefont {Feng}\ \emph
  {et~al.}(2017{\natexlab{a}})\citenamefont {Feng}, \citenamefont
  {El-Ganainy},\ and\ \citenamefont {Ge}}]{NHg6}%
  \BibitemOpen
  \bibfield  {author} {\bibinfo {author} {\bibfnamefont {L.}~\bibnamefont
  {Feng}}, \bibinfo {author} {\bibfnamefont {R.}~\bibnamefont {El-Ganainy}},\
  and\ \bibinfo {author} {\bibfnamefont {L.}~\bibnamefont {Ge}},\ }\bibfield
  {title} {\bibinfo {title} {Non-hermitian photonics based on parity-time
  symmetry},\ }\href {https://www.nature.com/articles/s41566-017-0031-1}
  {\bibfield  {journal} {\bibinfo  {journal} {Nat. Photon.}\ }\textbf {\bibinfo
  {volume} {11}},\ \bibinfo {pages} {752} (\bibinfo {year}
  {2017}{\natexlab{a}})}\BibitemShut {NoStop}%
\bibitem [{\citenamefont {El-Ganainy}\ \emph {et~al.}(2018)\citenamefont
  {El-Ganainy}, \citenamefont {Makris}, \citenamefont {Khajavikhan},
  \citenamefont {Musslimani}, \citenamefont {Rotter},\ and\ \citenamefont
  {Christodoulides}}]{NHg7}%
  \BibitemOpen
  \bibfield  {author} {\bibinfo {author} {\bibfnamefont {R.}~\bibnamefont
  {El-Ganainy}}, \bibinfo {author} {\bibfnamefont {K.~G.}\ \bibnamefont
  {Makris}}, \bibinfo {author} {\bibfnamefont {M.}~\bibnamefont {Khajavikhan}},
  \bibinfo {author} {\bibfnamefont {Z.~H.}\ \bibnamefont {Musslimani}},
  \bibinfo {author} {\bibfnamefont {S.}~\bibnamefont {Rotter}},\ and\ \bibinfo
  {author} {\bibfnamefont {D.~N.}\ \bibnamefont {Christodoulides}},\ }\bibfield
   {title} {\bibinfo {title} {Non-hermitian physics and {PT} symmetry},\ }\href
  {https://www.nature.com/articles/nphys4323} {\bibfield  {journal} {\bibinfo
  {journal} {Nat. Phys.}\ }\textbf {\bibinfo {volume} {14}},\ \bibinfo {pages}
  {11} (\bibinfo {year} {2018})}\BibitemShut {NoStop}%
\bibitem [{\citenamefont {Gong}\ \emph {et~al.}(2018)\citenamefont {Gong},
  \citenamefont {Ashida}, \citenamefont {Kawabata}, \citenamefont {Takasan},
  \citenamefont {Higashikawa},\ and\ \citenamefont {Ueda}}]{GongPRX}%
  \BibitemOpen
  \bibfield  {author} {\bibinfo {author} {\bibfnamefont {Z.}~\bibnamefont
  {Gong}}, \bibinfo {author} {\bibfnamefont {Y.}~\bibnamefont {Ashida}},
  \bibinfo {author} {\bibfnamefont {K.}~\bibnamefont {Kawabata}}, \bibinfo
  {author} {\bibfnamefont {K.}~\bibnamefont {Takasan}}, \bibinfo {author}
  {\bibfnamefont {S.}~\bibnamefont {Higashikawa}},\ and\ \bibinfo {author}
  {\bibfnamefont {M.}~\bibnamefont {Ueda}},\ }\bibfield  {title} {\bibinfo
  {title} {Topological {P}hases of {N}on-{H}ermitian {S}ystems},\ }\href
  {https://journals.aps.org/prx/abstract/10.1103/PhysRevX.8.031079} {\bibfield
  {journal} {\bibinfo  {journal} {Phys. Rev. X}\ }\textbf {\bibinfo {volume}
  {8}},\ \bibinfo {pages} {031079} (\bibinfo {year} {2018})}\BibitemShut
  {NoStop}%
\bibitem [{\citenamefont {Kawabata}\ \emph {et~al.}(2019)\citenamefont
  {Kawabata}, \citenamefont {Shiozaki}, \citenamefont {Ueda},\ and\
  \citenamefont {Sato}}]{NHg8}%
  \BibitemOpen
  \bibfield  {author} {\bibinfo {author} {\bibfnamefont {K.}~\bibnamefont
  {Kawabata}}, \bibinfo {author} {\bibfnamefont {K.}~\bibnamefont {Shiozaki}},
  \bibinfo {author} {\bibfnamefont {M.}~\bibnamefont {Ueda}},\ and\ \bibinfo
  {author} {\bibfnamefont {M.}~\bibnamefont {Sato}},\ }\bibfield  {title}
  {\bibinfo {title} {Symmetry and {T}opology in {N}on-{H}ermitian {P}hysics},\
  }\href {https://journals.aps.org/prx/abstract/10.1103/PhysRevX.9.041015}
  {\bibfield  {journal} {\bibinfo  {journal} {Phys. Rev. X}\ }\textbf {\bibinfo
  {volume} {9}},\ \bibinfo {pages} {041015} (\bibinfo {year}
  {2019})}\BibitemShut {NoStop}%
\bibitem [{\citenamefont {Heiss}(2012)}]{EpHeiss}%
  \BibitemOpen
  \bibfield  {author} {\bibinfo {author} {\bibfnamefont {W.~D.}\ \bibnamefont
  {Heiss}},\ }\bibfield  {title} {\bibinfo {title} {The physics of exceptional
  points},\ }\href
  {https://iopscience.iop.org/article/10.1088/1751-8113/45/44/444016}
  {\bibfield  {journal} {\bibinfo  {journal} {J. Phys. A}\ }\textbf {\bibinfo
  {volume} {45}},\ \bibinfo {pages} {444016} (\bibinfo {year}
  {2012})}\BibitemShut {NoStop}%
\bibitem [{\citenamefont {Minganti}\ \emph {et~al.}(2019)\citenamefont
  {Minganti}, \citenamefont {Miranowicz}, \citenamefont {Chhajlany},\ and\
  \citenamefont {Nori}}]{LiouvillianEp}%
  \BibitemOpen
  \bibfield  {author} {\bibinfo {author} {\bibfnamefont {F.}~\bibnamefont
  {Minganti}}, \bibinfo {author} {\bibfnamefont {A.}~\bibnamefont
  {Miranowicz}}, \bibinfo {author} {\bibfnamefont {R.~W.}\ \bibnamefont
  {Chhajlany}},\ and\ \bibinfo {author} {\bibfnamefont {F.}~\bibnamefont
  {Nori}},\ }\bibfield  {title} {\bibinfo {title} {Quantum exceptional points
  of non-{H}ermitian {H}amiltonians and {L}iouvillians: {T}he effects of
  quantum jumps},\ }\href
  {https://journals.aps.org/pra/abstract/10.1103/PhysRevA.100.062131}
  {\bibfield  {journal} {\bibinfo  {journal} {Phys. Rev. A}\ }\textbf {\bibinfo
  {volume} {100}},\ \bibinfo {pages} {062131} (\bibinfo {year}
  {2019})}\BibitemShut {NoStop}%
\bibitem [{\citenamefont {Miri}\ and\ \citenamefont {Al\`{u}}(2019)}]{EpAlu}%
  \BibitemOpen
  \bibfield  {author} {\bibinfo {author} {\bibfnamefont {M.-A.}\ \bibnamefont
  {Miri}}\ and\ \bibinfo {author} {\bibfnamefont {A.}~\bibnamefont {Al\`{u}}},\
  }\bibfield  {title} {\bibinfo {title} {Exceptional points in optics and
  photonics},\ }\href {https://www.science.org/doi/10.1126/science.aar7709}
  {\bibfield  {journal} {\bibinfo  {journal} {Science}\ }\textbf {\bibinfo
  {volume} {363}},\ \bibinfo {pages} {eaar7709} (\bibinfo {year}
  {2019})}\BibitemShut {NoStop}%
\bibitem [{\citenamefont {Bergholtz}\ \emph {et~al.}(2021)\citenamefont
  {Bergholtz}, \citenamefont {Budich},\ and\ \citenamefont {Kunst}}]{EpTopo}%
  \BibitemOpen
  \bibfield  {author} {\bibinfo {author} {\bibfnamefont {E.~J.}\ \bibnamefont
  {Bergholtz}}, \bibinfo {author} {\bibfnamefont {J.~C.}\ \bibnamefont
  {Budich}},\ and\ \bibinfo {author} {\bibfnamefont {F.~K.}\ \bibnamefont
  {Kunst}},\ }\bibfield  {title} {\bibinfo {title} {Exceptional topology of
  non-{H}ermitian systems},\ }\href
  {https://journals.aps.org/rmp/abstract/10.1103/RevModPhys.93.015005}
  {\bibfield  {journal} {\bibinfo  {journal} {Rev. Mod. Phys.}\ }\textbf
  {\bibinfo {volume} {93}},\ \bibinfo {pages} {015005} (\bibinfo {year}
  {2021})}\BibitemShut {NoStop}%
\bibitem [{\citenamefont {Brody}(2014)}]{biortho1}%
  \BibitemOpen
  \bibfield  {author} {\bibinfo {author} {\bibfnamefont {D.~C.}\ \bibnamefont
  {Brody}},\ }\bibfield  {title} {\bibinfo {title} {Biorthogonal quantum
  mechanics},\ }\href
  {https://iopscience.iop.org/article/10.1088/1751-8113/47/3/035305} {\bibfield
   {journal} {\bibinfo  {journal} {J. Phys. A}\ }\textbf {\bibinfo {volume}
  {47}},\ \bibinfo {pages} {035305} (\bibinfo {year} {2014})}\BibitemShut
  {NoStop}%
\bibitem [{\citenamefont {Kunst}\ \emph {et~al.}(2018)\citenamefont {Kunst},
  \citenamefont {Edvardsson}, \citenamefont {Budich},\ and\ \citenamefont
  {Bergholtz}}]{biortho2}%
  \BibitemOpen
  \bibfield  {author} {\bibinfo {author} {\bibfnamefont {F.~K.}\ \bibnamefont
  {Kunst}}, \bibinfo {author} {\bibfnamefont {E.}~\bibnamefont {Edvardsson}},
  \bibinfo {author} {\bibfnamefont {J.~C.}\ \bibnamefont {Budich}},\ and\
  \bibinfo {author} {\bibfnamefont {E.~J.}\ \bibnamefont {Bergholtz}},\
  }\bibfield  {title} {\bibinfo {title} {Biorthogonal {B}ulk-{B}oundary
  {C}orrespondence in {N}on-{H}ermitian {S}ystems},\ }\href
  {https://journals.aps.org/prl/abstract/10.1103/PhysRevLett.121.026808}
  {\bibfield  {journal} {\bibinfo  {journal} {Phys. Rev. Lett.}\ }\textbf
  {\bibinfo {volume} {121}},\ \bibinfo {pages} {026808} (\bibinfo {year}
  {2018})}\BibitemShut {NoStop}%
\bibitem [{\citenamefont {Yokomizo}\ and\ \citenamefont
  {Murakami}(2019)}]{NB1}%
  \BibitemOpen
  \bibfield  {author} {\bibinfo {author} {\bibfnamefont {K.}~\bibnamefont
  {Yokomizo}}\ and\ \bibinfo {author} {\bibfnamefont {S.}~\bibnamefont
  {Murakami}},\ }\bibfield  {title} {\bibinfo {title} {Non-{B}loch {B}and
  {T}heory of {N}on-{H}ermitian {S}ystems},\ }\href
  {https://journals.aps.org/prl/abstract/10.1103/PhysRevLett.123.066404}
  {\bibfield  {journal} {\bibinfo  {journal} {Phys. Rev. Lett.}\ }\textbf
  {\bibinfo {volume} {123}},\ \bibinfo {pages} {066404} (\bibinfo {year}
  {2019})}\BibitemShut {NoStop}%
\bibitem [{\citenamefont {Yokomizo}\ and\ \citenamefont
  {Murakami}(2020)}]{NB2}%
  \BibitemOpen
  \bibfield  {author} {\bibinfo {author} {\bibfnamefont {K.}~\bibnamefont
  {Yokomizo}}\ and\ \bibinfo {author} {\bibfnamefont {S.}~\bibnamefont
  {Murakami}},\ }\bibfield  {title} {\bibinfo {title} {Non-{B}loch band theory
  and bulk–edge correspondence in non-{H}ermitian systems},\ }\href
  {https://academic.oup.com/ptep/article/2020/12/12A102/5906038} {\bibfield
  {journal} {\bibinfo  {journal} {Prog. Theor. Exp. Phys.}\ }\textbf {\bibinfo
  {volume} {2020}},\ \bibinfo {pages} {12A102} (\bibinfo {year}
  {2020})}\BibitemShut {NoStop}%
\bibitem [{\citenamefont {Kawabata}\ \emph
  {et~al.}(2020{\natexlab{a}})\citenamefont {Kawabata}, \citenamefont {Okuma},\
  and\ \citenamefont {Sato}}]{NB3}%
  \BibitemOpen
  \bibfield  {author} {\bibinfo {author} {\bibfnamefont {K.}~\bibnamefont
  {Kawabata}}, \bibinfo {author} {\bibfnamefont {N.}~\bibnamefont {Okuma}},\
  and\ \bibinfo {author} {\bibfnamefont {M.}~\bibnamefont {Sato}},\ }\bibfield
  {title} {\bibinfo {title} {Non-{B}loch band theory of non-{H}ermitian
  {H}amiltonians in the symplectic class},\ }\href
  {https://journals.aps.org/prb/abstract/10.1103/PhysRevB.101.195147}
  {\bibfield  {journal} {\bibinfo  {journal} {Phys. Rev. B}\ }\textbf {\bibinfo
  {volume} {101}},\ \bibinfo {pages} {195147} (\bibinfo {year}
  {2020}{\natexlab{a}})}\BibitemShut {NoStop}%
\bibitem [{\citenamefont {Xue}\ \emph {et~al.}(2021)\citenamefont {Xue},
  \citenamefont {Li}, \citenamefont {Hu}, \citenamefont {Song},\ and\
  \citenamefont {Wang}}]{NB4}%
  \BibitemOpen
  \bibfield  {author} {\bibinfo {author} {\bibfnamefont {W.-T.}\ \bibnamefont
  {Xue}}, \bibinfo {author} {\bibfnamefont {M.-R.}\ \bibnamefont {Li}},
  \bibinfo {author} {\bibfnamefont {Y.-M.}\ \bibnamefont {Hu}}, \bibinfo
  {author} {\bibfnamefont {F.}~\bibnamefont {Song}},\ and\ \bibinfo {author}
  {\bibfnamefont {Z.}~\bibnamefont {Wang}},\ }\bibfield  {title} {\bibinfo
  {title} {Simple formulas of directional amplification from non-{B}loch band
  theory},\ }\href
  {https://journals.aps.org/prb/abstract/10.1103/PhysRevB.103.L241408}
  {\bibfield  {journal} {\bibinfo  {journal} {Phys. Rev. B}\ }\textbf {\bibinfo
  {volume} {103}},\ \bibinfo {pages} {L241408} (\bibinfo {year}
  {2021})}\BibitemShut {NoStop}%
\bibitem [{\citenamefont {Yokomizo}\ and\ \citenamefont
  {Murakami}(2023)}]{NB5}%
  \BibitemOpen
  \bibfield  {author} {\bibinfo {author} {\bibfnamefont {K.}~\bibnamefont
  {Yokomizo}}\ and\ \bibinfo {author} {\bibfnamefont {S.}~\bibnamefont
  {Murakami}},\ }\bibfield  {title} {\bibinfo {title} {Non-{B}loch bands in
  two-dimensional non-{H}ermitian systems},\ }\href
  {https://journals.aps.org/prb/abstract/10.1103/PhysRevB.107.195112}
  {\bibfield  {journal} {\bibinfo  {journal} {Phys. Rev. B}\ }\textbf {\bibinfo
  {volume} {107}},\ \bibinfo {pages} {195112} (\bibinfo {year}
  {2023})}\BibitemShut {NoStop}%
\bibitem [{\citenamefont {Hatano}\ and\ \citenamefont
  {Nelson}(1996)}]{HNmodel}%
  \BibitemOpen
  \bibfield  {author} {\bibinfo {author} {\bibfnamefont {N.}~\bibnamefont
  {Hatano}}\ and\ \bibinfo {author} {\bibfnamefont {D.~R.}\ \bibnamefont
  {Nelson}},\ }\bibfield  {title} {\bibinfo {title} {Localization {T}ransitions
  in {N}on-{H}ermitian {Q}uantum {M}echanics},\ }\href
  {https://journals.aps.org/prl/abstract/10.1103/PhysRevLett.77.570} {\bibfield
   {journal} {\bibinfo  {journal} {Phys. Rev. Lett.}\ }\textbf {\bibinfo
  {volume} {77}},\ \bibinfo {pages} {570} (\bibinfo {year} {1996})}\BibitemShut
  {NoStop}%
\bibitem [{\citenamefont {Longhi}\ \emph
  {et~al.}(2015{\natexlab{a}})\citenamefont {Longhi}, \citenamefont {Gatti},\
  and\ \citenamefont {Valle}}]{LonghiSR}%
  \BibitemOpen
  \bibfield  {author} {\bibinfo {author} {\bibfnamefont {S.}~\bibnamefont
  {Longhi}}, \bibinfo {author} {\bibfnamefont {D.}~\bibnamefont {Gatti}},\ and\
  \bibinfo {author} {\bibfnamefont {G.~D.}\ \bibnamefont {Valle}},\ }\bibfield
  {title} {\bibinfo {title} {Robust light transport in non-{H}ermitian photonic
  lattices},\ }\href {https://www.nature.com/articles/srep13376} {\bibfield
  {journal} {\bibinfo  {journal} {Sci. Rep.}\ }\textbf {\bibinfo {volume}
  {5}},\ \bibinfo {pages} {13376} (\bibinfo {year}
  {2015}{\natexlab{a}})}\BibitemShut {NoStop}%
\bibitem [{\citenamefont {Longhi}\ \emph
  {et~al.}(2015{\natexlab{b}})\citenamefont {Longhi}, \citenamefont {Gatti},\
  and\ \citenamefont {Valle}}]{Longhi2015prb}%
  \BibitemOpen
  \bibfield  {author} {\bibinfo {author} {\bibfnamefont {S.}~\bibnamefont
  {Longhi}}, \bibinfo {author} {\bibfnamefont {D.}~\bibnamefont {Gatti}},\ and\
  \bibinfo {author} {\bibfnamefont {G.~D.}\ \bibnamefont {Valle}},\ }\bibfield
  {title} {\bibinfo {title} {Non-{H}ermitian transparency and one-way transport
  in low-dimensional lattices by an imaginary gauge field},\ }\href
  {https://journals.aps.org/prb/abstract/10.1103/PhysRevB.92.094204} {\bibfield
   {journal} {\bibinfo  {journal} {Phys. Rev. B}\ }\textbf {\bibinfo {volume}
  {92}},\ \bibinfo {pages} {094204} (\bibinfo {year}
  {2015}{\natexlab{b}})}\BibitemShut {NoStop}%
\bibitem [{\citenamefont {Metelmann}\ and\ \citenamefont
  {Clerk}(2015)}]{AnjaPRX}%
  \BibitemOpen
  \bibfield  {author} {\bibinfo {author} {\bibfnamefont {A.}~\bibnamefont
  {Metelmann}}\ and\ \bibinfo {author} {\bibfnamefont {A.~A.}\ \bibnamefont
  {Clerk}},\ }\bibfield  {title} {\bibinfo {title} {Nonreciprocal {P}hoton
  {T}ransmission and {A}mplification via {R}eservoir {E}ngineering},\ }\href
  {https://journals.aps.org/prx/abstract/10.1103/PhysRevX.5.021025} {\bibfield
  {journal} {\bibinfo  {journal} {Phys. Rev. X}\ }\textbf {\bibinfo {volume}
  {5}},\ \bibinfo {pages} {021025} (\bibinfo {year} {2015})}\BibitemShut
  {NoStop}%
\bibitem [{\citenamefont {Weidemann}\ \emph {et~al.}(2020)\citenamefont
  {Weidemann}, \citenamefont {Kremer}, \citenamefont {Helbig}, \citenamefont
  {Hofmann}, \citenamefont {Stegmaier}, \citenamefont {Greiter}, \citenamefont
  {Thomale},\ and\ \citenamefont {Szameit}}]{Qwalk1}%
  \BibitemOpen
  \bibfield  {author} {\bibinfo {author} {\bibfnamefont {S.}~\bibnamefont
  {Weidemann}}, \bibinfo {author} {\bibfnamefont {M.}~\bibnamefont {Kremer}},
  \bibinfo {author} {\bibfnamefont {T.}~\bibnamefont {Helbig}}, \bibinfo
  {author} {\bibfnamefont {T.}~\bibnamefont {Hofmann}}, \bibinfo {author}
  {\bibfnamefont {A.}~\bibnamefont {Stegmaier}}, \bibinfo {author}
  {\bibfnamefont {M.}~\bibnamefont {Greiter}}, \bibinfo {author} {\bibfnamefont
  {R.}~\bibnamefont {Thomale}},\ and\ \bibinfo {author} {\bibfnamefont
  {A.}~\bibnamefont {Szameit}},\ }\bibfield  {title} {\bibinfo {title}
  {Topological funneling of light},\ }\href
  {https://www.science.org/doi/10.1126/science.aaz8727} {\bibfield  {journal}
  {\bibinfo  {journal} {Science}\ }\textbf {\bibinfo {volume} {368}},\ \bibinfo
  {pages} {311} (\bibinfo {year} {2020})}\BibitemShut {NoStop}%
\bibitem [{\citenamefont {Weidemann}\ \emph {et~al.}(2022)\citenamefont
  {Weidemann}, \citenamefont {Kremer}, \citenamefont {Longhi},\ and\
  \citenamefont {Szameit}}]{Qwalk2}%
  \BibitemOpen
  \bibfield  {author} {\bibinfo {author} {\bibfnamefont {S.}~\bibnamefont
  {Weidemann}}, \bibinfo {author} {\bibfnamefont {M.}~\bibnamefont {Kremer}},
  \bibinfo {author} {\bibfnamefont {S.}~\bibnamefont {Longhi}},\ and\ \bibinfo
  {author} {\bibfnamefont {A.}~\bibnamefont {Szameit}},\ }\bibfield  {title}
  {\bibinfo {title} {Topological triple phase transition in non-{H}ermitian
  {F}loquet quasicrystals},\ }\href
  {https://www.nature.com/articles/s41586-021-04253-0} {\bibfield  {journal}
  {\bibinfo  {journal} {Nature (London)}\ }\textbf {\bibinfo {volume} {601}},\
  \bibinfo {pages} {354} (\bibinfo {year} {2022})}\BibitemShut {NoStop}%
\bibitem [{\citenamefont {Liang}\ \emph {et~al.}(2022)\citenamefont {Liang},
  \citenamefont {Xie}, \citenamefont {Dong}, \citenamefont {Li}, \citenamefont
  {Li}, \citenamefont {Gadway}, \citenamefont {Yi},\ and\ \citenamefont
  {Yan}}]{Qwalk3}%
  \BibitemOpen
  \bibfield  {author} {\bibinfo {author} {\bibfnamefont {Q.}~\bibnamefont
  {Liang}}, \bibinfo {author} {\bibfnamefont {D.}~\bibnamefont {Xie}}, \bibinfo
  {author} {\bibfnamefont {Z.}~\bibnamefont {Dong}}, \bibinfo {author}
  {\bibfnamefont {H.}~\bibnamefont {Li}}, \bibinfo {author} {\bibfnamefont
  {H.}~\bibnamefont {Li}}, \bibinfo {author} {\bibfnamefont {B.}~\bibnamefont
  {Gadway}}, \bibinfo {author} {\bibfnamefont {W.}~\bibnamefont {Yi}},\ and\
  \bibinfo {author} {\bibfnamefont {B.}~\bibnamefont {Yan}},\ }\bibfield
  {title} {\bibinfo {title} {Dynamic {S}ignatures of {N}on-{H}ermitian {S}kin
  {E}ffect and {T}opology in {U}ltracold {A}toms},\ }\href
  {https://journals.aps.org/prl/abstract/10.1103/PhysRevLett.129.070401}
  {\bibfield  {journal} {\bibinfo  {journal} {Phys. Rev. Lett.}\ }\textbf
  {\bibinfo {volume} {129}},\ \bibinfo {pages} {070401} (\bibinfo {year}
  {2022})}\BibitemShut {NoStop}%
\bibitem [{\citenamefont {Song}\ \emph {et~al.}(2020)\citenamefont {Song},
  \citenamefont {Liu}, \citenamefont {Zheng}, \citenamefont {Zhang},
  \citenamefont {Wang},\ and\ \citenamefont {Lu}}]{LuSkin2D}%
  \BibitemOpen
  \bibfield  {author} {\bibinfo {author} {\bibfnamefont {Y.}~\bibnamefont
  {Song}}, \bibinfo {author} {\bibfnamefont {W.}~\bibnamefont {Liu}}, \bibinfo
  {author} {\bibfnamefont {L.}~\bibnamefont {Zheng}}, \bibinfo {author}
  {\bibfnamefont {Y.}~\bibnamefont {Zhang}}, \bibinfo {author} {\bibfnamefont
  {B.}~\bibnamefont {Wang}},\ and\ \bibinfo {author} {\bibfnamefont
  {P.}~\bibnamefont {Lu}},\ }\bibfield  {title} {\bibinfo {title}
  {Two-dimensional non-{H}ermitian {S}kin {E}ffect in a {S}ynthetic {P}hotonic
  {L}attice},\ }\href
  {https://journals.aps.org/prapplied/abstract/10.1103/PhysRevApplied.14.064076}
  {\bibfield  {journal} {\bibinfo  {journal} {Phys. Rev. Appl.}\ }\textbf
  {\bibinfo {volume} {14}},\ \bibinfo {pages} {064076} (\bibinfo {year}
  {2020})}\BibitemShut {NoStop}%
\bibitem [{\citenamefont {Wang}\ \emph {et~al.}(2021)\citenamefont {Wang},
  \citenamefont {Dutt}, \citenamefont {Yang}, \citenamefont {Wojcik},
  \citenamefont {Vu\v{c}kovi\'{c}},\ and\ \citenamefont {Fan}}]{FanScience21}%
  \BibitemOpen
  \bibfield  {author} {\bibinfo {author} {\bibfnamefont {K.}~\bibnamefont
  {Wang}}, \bibinfo {author} {\bibfnamefont {A.}~\bibnamefont {Dutt}}, \bibinfo
  {author} {\bibfnamefont {K.~Y.}\ \bibnamefont {Yang}}, \bibinfo {author}
  {\bibfnamefont {C.~C.}\ \bibnamefont {Wojcik}}, \bibinfo {author}
  {\bibfnamefont {J.}~\bibnamefont {Vu\v{c}kovi\'{c}}},\ and\ \bibinfo {author}
  {\bibfnamefont {S.}~\bibnamefont {Fan}},\ }\bibfield  {title} {\bibinfo
  {title} {Generating arbitrary topological windings of a non-{H}ermitian
  band},\ }\href {https://www.science.org/doi/10.1126/science.abf6568}
  {\bibfield  {journal} {\bibinfo  {journal} {Science}\ }\textbf {\bibinfo
  {volume} {371}},\ \bibinfo {pages} {1240} (\bibinfo {year}
  {2021})}\BibitemShut {NoStop}%
\bibitem [{\citenamefont {Yao}\ and\ \citenamefont {Wang}(2018)}]{Yao2018}%
  \BibitemOpen
  \bibfield  {author} {\bibinfo {author} {\bibfnamefont {S.}~\bibnamefont
  {Yao}}\ and\ \bibinfo {author} {\bibfnamefont {Z.}~\bibnamefont {Wang}},\
  }\bibfield  {title} {\bibinfo {title} {Edge {S}tates and {T}opological
  {I}nvariants of {N}on-{H}ermitian {S}ystems},\ }\href
  {https://journals.aps.org/prl/abstract/10.1103/PhysRevLett.121.086803}
  {\bibfield  {journal} {\bibinfo  {journal} {Phys. Rev. Lett.}\ }\textbf
  {\bibinfo {volume} {121}},\ \bibinfo {pages} {086803} (\bibinfo {year}
  {2018})}\BibitemShut {NoStop}%
\bibitem [{\citenamefont {Yao}\ \emph {et~al.}(2018)\citenamefont {Yao},
  \citenamefont {Song},\ and\ \citenamefont {Wang}}]{Yao20182}%
  \BibitemOpen
  \bibfield  {author} {\bibinfo {author} {\bibfnamefont {S.}~\bibnamefont
  {Yao}}, \bibinfo {author} {\bibfnamefont {F.}~\bibnamefont {Song}},\ and\
  \bibinfo {author} {\bibfnamefont {Z.}~\bibnamefont {Wang}},\ }\bibfield
  {title} {\bibinfo {title} {Non-{H}ermitian {C}hern {B}ands},\ }\href
  {https://journals.aps.org/prl/abstract/10.1103/PhysRevLett.121.136802}
  {\bibfield  {journal} {\bibinfo  {journal} {Phys. Rev. Lett.}\ }\textbf
  {\bibinfo {volume} {121}},\ \bibinfo {pages} {136802} (\bibinfo {year}
  {2018})}\BibitemShut {NoStop}%
\bibitem [{\citenamefont {Okuma}\ \emph {et~al.}(2020)\citenamefont {Okuma},
  \citenamefont {Kawabata}, \citenamefont {Shiozaki},\ and\ \citenamefont
  {Sato}}]{TopoOrigin}%
  \BibitemOpen
  \bibfield  {author} {\bibinfo {author} {\bibfnamefont {N.}~\bibnamefont
  {Okuma}}, \bibinfo {author} {\bibfnamefont {K.}~\bibnamefont {Kawabata}},
  \bibinfo {author} {\bibfnamefont {K.}~\bibnamefont {Shiozaki}},\ and\
  \bibinfo {author} {\bibfnamefont {M.}~\bibnamefont {Sato}},\ }\bibfield
  {title} {\bibinfo {title} {Topological {O}rigin of {N}on-{H}ermitian {S}kin
  {E}ffects},\ }\href
  {https://journals.aps.org/prl/abstract/10.1103/PhysRevLett.124.086801}
  {\bibfield  {journal} {\bibinfo  {journal} {Phys. Rev. Lett.}\ }\textbf
  {\bibinfo {volume} {124}},\ \bibinfo {pages} {086801} (\bibinfo {year}
  {2020})}\BibitemShut {NoStop}%
\bibitem [{\citenamefont {Zhu}\ \emph {et~al.}(2020)\citenamefont {Zhu},
  \citenamefont {Wang}, \citenamefont {Gupta}, \citenamefont {Zhang},
  \citenamefont {Xie}, \citenamefont {Lu},\ and\ \citenamefont
  {Chen}}]{YFChen2020}%
  \BibitemOpen
  \bibfield  {author} {\bibinfo {author} {\bibfnamefont {X.}~\bibnamefont
  {Zhu}}, \bibinfo {author} {\bibfnamefont {H.}~\bibnamefont {Wang}}, \bibinfo
  {author} {\bibfnamefont {S.~K.}\ \bibnamefont {Gupta}}, \bibinfo {author}
  {\bibfnamefont {H.}~\bibnamefont {Zhang}}, \bibinfo {author} {\bibfnamefont
  {B.}~\bibnamefont {Xie}}, \bibinfo {author} {\bibfnamefont {M.}~\bibnamefont
  {Lu}},\ and\ \bibinfo {author} {\bibfnamefont {Y.}~\bibnamefont {Chen}},\
  }\bibfield  {title} {\bibinfo {title} {Photonic non-{H}ermitian skin effect
  and non-{B}loch bulk-boundary correspondence},\ }\href
  {https://journals.aps.org/prresearch/abstract/10.1103/PhysRevResearch.2.013280}
  {\bibfield  {journal} {\bibinfo  {journal} {Phys. Rev. Res.}\ }\textbf
  {\bibinfo {volume} {2}},\ \bibinfo {pages} {013280} (\bibinfo {year}
  {2020})}\BibitemShut {NoStop}%
\bibitem [{\citenamefont {Budich}\ and\ \citenamefont
  {Bergholtz}(2020)}]{TopoSensor}%
  \BibitemOpen
  \bibfield  {author} {\bibinfo {author} {\bibfnamefont {J.~C.}\ \bibnamefont
  {Budich}}\ and\ \bibinfo {author} {\bibfnamefont {E.~J.}\ \bibnamefont
  {Bergholtz}},\ }\bibfield  {title} {\bibinfo {title} {Non-{H}ermitian
  {T}opological {S}ensors},\ }\href
  {https://journals.aps.org/prl/abstract/10.1103/PhysRevLett.125.180403}
  {\bibfield  {journal} {\bibinfo  {journal} {Phys. Rev. Lett.}\ }\textbf
  {\bibinfo {volume} {125}},\ \bibinfo {pages} {180403} (\bibinfo {year}
  {2020})}\BibitemShut {NoStop}%
\bibitem [{\citenamefont {Kawabata}\ \emph
  {et~al.}(2020{\natexlab{b}})\citenamefont {Kawabata}, \citenamefont {Sato},\
  and\ \citenamefont {Shiozaki}}]{HigherOrderSkin}%
  \BibitemOpen
  \bibfield  {author} {\bibinfo {author} {\bibfnamefont {K.}~\bibnamefont
  {Kawabata}}, \bibinfo {author} {\bibfnamefont {M.}~\bibnamefont {Sato}},\
  and\ \bibinfo {author} {\bibfnamefont {K.}~\bibnamefont {Shiozaki}},\
  }\bibfield  {title} {\bibinfo {title} {Higher-order non-{H}ermitian skin
  effect},\ }\href
  {https://journals.aps.org/prb/abstract/10.1103/PhysRevB.102.205118}
  {\bibfield  {journal} {\bibinfo  {journal} {Phys. Rev. B}\ }\textbf {\bibinfo
  {volume} {102}},\ \bibinfo {pages} {205118} (\bibinfo {year}
  {2020}{\natexlab{b}})}\BibitemShut {NoStop}%
\bibitem [{\citenamefont {Longhi}(2016)}]{LonghiQD}%
  \BibitemOpen
  \bibfield  {author} {\bibinfo {author} {\bibfnamefont {S.}~\bibnamefont
  {Longhi}},\ }\bibfield  {title} {\bibinfo {title} {Quantum decay and
  amplification in a non-{H}ermitian unstable continuum},\ }\href
  {https://journals.aps.org/pra/abstract/10.1103/PhysRevA.93.062129} {\bibfield
   {journal} {\bibinfo  {journal} {Phys. Rev. A}\ }\textbf {\bibinfo {volume}
  {93}},\ \bibinfo {pages} {062129} (\bibinfo {year} {2016})}\BibitemShut
  {NoStop}%
\bibitem [{\citenamefont {Roccati}\ \emph {et~al.}(2022)\citenamefont
  {Roccati}, \citenamefont {Lorenzo}, \citenamefont {Calaj\`{o}}, \citenamefont
  {Palma}, \citenamefont {Carollo},\ and\ \citenamefont
  {Ciccarello}}]{Federico}%
  \BibitemOpen
  \bibfield  {author} {\bibinfo {author} {\bibfnamefont {F.}~\bibnamefont
  {Roccati}}, \bibinfo {author} {\bibfnamefont {S.}~\bibnamefont {Lorenzo}},
  \bibinfo {author} {\bibfnamefont {G.}~\bibnamefont {Calaj\`{o}}}, \bibinfo
  {author} {\bibfnamefont {G.~M.}\ \bibnamefont {Palma}}, \bibinfo {author}
  {\bibfnamefont {A.}~\bibnamefont {Carollo}},\ and\ \bibinfo {author}
  {\bibfnamefont {F.}~\bibnamefont {Ciccarello}},\ }\bibfield  {title}
  {\bibinfo {title} {Exotic interactions mediated by a non-{H}ermitian photonic
  bath},\ }\href
  {https://opg.optica.org/optica/fulltext.cfm?uri=optica-9-5-565&id=472904}
  {\bibfield  {journal} {\bibinfo  {journal} {Optica}\ }\textbf {\bibinfo
  {volume} {9}},\ \bibinfo {pages} {565} (\bibinfo {year} {2022})}\BibitemShut
  {NoStop}%
\bibitem [{\citenamefont {Gong}\ \emph
  {et~al.}(2023{\natexlab{a}})\citenamefont {Gong}, \citenamefont {Bello},
  \citenamefont {Malz},\ and\ \citenamefont {Kunst}}]{GongHN}%
  \BibitemOpen
  \bibfield  {author} {\bibinfo {author} {\bibfnamefont {Z.}~\bibnamefont
  {Gong}}, \bibinfo {author} {\bibfnamefont {M.}~\bibnamefont {Bello}},
  \bibinfo {author} {\bibfnamefont {D.}~\bibnamefont {Malz}},\ and\ \bibinfo
  {author} {\bibfnamefont {F.~K.}\ \bibnamefont {Kunst}},\ }\bibfield  {title}
  {\bibinfo {title} {Anomalous {B}ehaviors of {Q}uantum {E}mitters in
  {N}on-{H}ermitian {B}aths},\ }\href
  {https://journals.aps.org/prl/abstract/10.1103/PhysRevLett.129.223601}
  {\bibfield  {journal} {\bibinfo  {journal} {Phys. Rev. Lett.}\ }\textbf
  {\bibinfo {volume} {129}},\ \bibinfo {pages} {223601} (\bibinfo {year}
  {2023}{\natexlab{a}})}\BibitemShut {NoStop}%
\bibitem [{\citenamefont {Gong}\ \emph
  {et~al.}(2023{\natexlab{b}})\citenamefont {Gong}, \citenamefont {Bello},
  \citenamefont {Malz},\ and\ \citenamefont {Kunst}}]{GongHN2}%
  \BibitemOpen
  \bibfield  {author} {\bibinfo {author} {\bibfnamefont {Z.}~\bibnamefont
  {Gong}}, \bibinfo {author} {\bibfnamefont {M.}~\bibnamefont {Bello}},
  \bibinfo {author} {\bibfnamefont {D.}~\bibnamefont {Malz}},\ and\ \bibinfo
  {author} {\bibfnamefont {F.~K.}\ \bibnamefont {Kunst}},\ }\bibfield  {title}
  {\bibinfo {title} {Bound states and photon emission in non-{H}ermitian
  nanophotonics},\ }\href
  {https://journals.aps.org/pra/abstract/10.1103/PhysRevA.106.053517}
  {\bibfield  {journal} {\bibinfo  {journal} {Phys. Rev. A}\ }\textbf {\bibinfo
  {volume} {106}},\ \bibinfo {pages} {053517} (\bibinfo {year}
  {2023}{\natexlab{b}})}\BibitemShut {NoStop}%
\bibitem [{\citenamefont {Lv}\ \emph {et~al.}(2022)\citenamefont {Lv},
  \citenamefont {Zhang}, \citenamefont {Zhai},\ and\ \citenamefont
  {Zhou}}]{curveLv}%
  \BibitemOpen
  \bibfield  {author} {\bibinfo {author} {\bibfnamefont {C.}~\bibnamefont
  {Lv}}, \bibinfo {author} {\bibfnamefont {R.}~\bibnamefont {Zhang}}, \bibinfo
  {author} {\bibfnamefont {Z.}~\bibnamefont {Zhai}},\ and\ \bibinfo {author}
  {\bibfnamefont {Q.}~\bibnamefont {Zhou}},\ }\bibfield  {title} {\bibinfo
  {title} {Curving the space by non-hermiticity},\ }\href
  {https://www.nature.com/articles/s41467-022-29774-8} {\bibfield  {journal}
  {\bibinfo  {journal} {Nat. Commun.}\ }\textbf {\bibinfo {volume} {13}},\
  \bibinfo {pages} {2184} (\bibinfo {year} {2022})}\BibitemShut {NoStop}%
\bibitem [{\citenamefont {Wang}\ \emph {et~al.}(2022)\citenamefont {Wang},
  \citenamefont {Wang}, \citenamefont {Yao}, \citenamefont {Shen},
  \citenamefont {Wu}, \citenamefont {Qian}, \citenamefont {Li}, \citenamefont
  {Zhu},\ and\ \citenamefont {You}}]{YouNC}%
  \BibitemOpen
  \bibfield  {author} {\bibinfo {author} {\bibfnamefont {Z.-Q.}\ \bibnamefont
  {Wang}}, \bibinfo {author} {\bibfnamefont {Y.-P.}\ \bibnamefont {Wang}},
  \bibinfo {author} {\bibfnamefont {J.}~\bibnamefont {Yao}}, \bibinfo {author}
  {\bibfnamefont {R.-C.}\ \bibnamefont {Shen}}, \bibinfo {author}
  {\bibfnamefont {W.-J.}\ \bibnamefont {Wu}}, \bibinfo {author} {\bibfnamefont
  {J.}~\bibnamefont {Qian}}, \bibinfo {author} {\bibfnamefont {J.}~\bibnamefont
  {Li}}, \bibinfo {author} {\bibfnamefont {S.-Y.}\ \bibnamefont {Zhu}},\ and\
  \bibinfo {author} {\bibfnamefont {J.~Q.}\ \bibnamefont {You}},\ }\bibfield
  {title} {\bibinfo {title} {Giant spin ensembles in waveguide magnonics},\
  }\href {https://www.nature.com/articles/s41467-022-35174-9} {\bibfield
  {journal} {\bibinfo  {journal} {Nat. Commun.}\ }\textbf {\bibinfo {volume}
  {13}},\ \bibinfo {pages} {7580} (\bibinfo {year} {2022})}\BibitemShut
  {NoStop}%
\bibitem [{\citenamefont {Longhi}(2020)}]{LonghiGA}%
  \BibitemOpen
  \bibfield  {author} {\bibinfo {author} {\bibfnamefont {S.}~\bibnamefont
  {Longhi}},\ }\bibfield  {title} {\bibinfo {title} {Photonic simulation of
  giant atom decay},\ }\href
  {https://opg.optica.org/ol/abstract.cfm?uri=ol-45-11-3017&origin=search}
  {\bibfield  {journal} {\bibinfo  {journal} {Opt. Lett.}\ }\textbf {\bibinfo
  {volume} {45}},\ \bibinfo {pages} {3017} (\bibinfo {year}
  {2020})}\BibitemShut {NoStop}%
\bibitem [{\citenamefont {Du}\ \emph {et~al.}(2022{\natexlab{b}})\citenamefont
  {Du}, \citenamefont {Chen}, \citenamefont {Zhang},\ and\ \citenamefont
  {Li}}]{DLprr2}%
  \BibitemOpen
  \bibfield  {author} {\bibinfo {author} {\bibfnamefont {L.}~\bibnamefont
  {Du}}, \bibinfo {author} {\bibfnamefont {Y.-T.}\ \bibnamefont {Chen}},
  \bibinfo {author} {\bibfnamefont {Y.}~\bibnamefont {Zhang}},\ and\ \bibinfo
  {author} {\bibfnamefont {Y.}~\bibnamefont {Li}},\ }\bibfield  {title}
  {\bibinfo {title} {Giant atoms with time-dependent couplings},\ }\href
  {https://journals.aps.org/prresearch/abstract/10.1103/PhysRevResearch.4.023198}
  {\bibfield  {journal} {\bibinfo  {journal} {Phys. Rev. Res.}\ }\textbf
  {\bibinfo {volume} {4}},\ \bibinfo {pages} {023198} (\bibinfo {year}
  {2022}{\natexlab{b}})}\BibitemShut {NoStop}%
\bibitem [{\citenamefont {Soro}\ \emph {et~al.}(2023)\citenamefont {Soro},
  \citenamefont {Mu\~{n}oz},\ and\ \citenamefont {Kockum}}]{AFKstructured}%
  \BibitemOpen
  \bibfield  {author} {\bibinfo {author} {\bibfnamefont {A.}~\bibnamefont
  {Soro}}, \bibinfo {author} {\bibfnamefont {C.~S.}\ \bibnamefont
  {Mu\~{n}oz}},\ and\ \bibinfo {author} {\bibfnamefont {A.~F.}\ \bibnamefont
  {Kockum}},\ }\bibfield  {title} {\bibinfo {title} {Interaction between giant
  atoms in a one-dimensional structured environment},\ }\href
  {https://journals.aps.org/pra/abstract/10.1103/PhysRevA.107.013710}
  {\bibfield  {journal} {\bibinfo  {journal} {Phys. Rev. A}\ }\textbf {\bibinfo
  {volume} {107}},\ \bibinfo {pages} {013710} (\bibinfo {year}
  {2023})}\BibitemShut {NoStop}%
\bibitem [{\citenamefont {Cohen-Tannoudji}\ \emph {et~al.}(1998)\citenamefont
  {Cohen-Tannoudji}, \citenamefont {Dupont-Roc},\ and\ \citenamefont
  {Grynberg}}]{resolvent}%
  \BibitemOpen
  \bibfield  {author} {\bibinfo {author} {\bibfnamefont {C.}~\bibnamefont
  {Cohen-Tannoudji}}, \bibinfo {author} {\bibfnamefont {J.}~\bibnamefont
  {Dupont-Roc}},\ and\ \bibinfo {author} {\bibfnamefont {G.}~\bibnamefont
  {Grynberg}},\ }\bibfield  {title} {\bibinfo {title} {Nonperturbative
  {C}alculation of {T}ransition {A}mplitudes},\ }in\ \href
  {https://onlinelibrary.wiley.com/doi/abs/10.1002/9783527617197.ch3} {\emph
  {\bibinfo {booktitle} {Atom—Photon Interactions}}}\ (\bibinfo  {publisher}
  {John Wiley \& Sons, Ltd},\ \bibinfo {year} {1998})\ Chap.~\bibinfo {chapter}
  {3}, pp.\ \bibinfo {pages} {165--255}\BibitemShut {NoStop}%
\bibitem [{SM()}]{SM}%
  \BibitemOpen
  \href@noop {} {\bibinfo  {journal} {See {S}upplemental {M}aterial at
  http://xxxxxx for more details}\ }\BibitemShut {NoStop}%
\bibitem [{\citenamefont {Sakurai}\ and\ \citenamefont
  {Napolitano}(2011)}]{WWapprox}%
  \BibitemOpen
\bibfield  {journal} {  }\bibfield  {author} {\bibinfo {author} {\bibfnamefont
  {J.~J.}\ \bibnamefont {Sakurai}}\ and\ \bibinfo {author} {\bibfnamefont
  {J.}~\bibnamefont {Napolitano}},\ }\href@noop {} {\emph {\bibinfo {title}
  {Modern Quantum Mechanics}}}\ (\bibinfo  {publisher} {Addison-Wesley,
  Boston},\ \bibinfo {year} {2011})\BibitemShut {NoStop}%
\bibitem [{\citenamefont {Gonz\'{a}lez-Tudela}\ and\ \citenamefont
  {Cirac}(2017{\natexlab{a}})}]{AGT2017prl}%
  \BibitemOpen
  \bibfield  {author} {\bibinfo {author} {\bibfnamefont {A.}~\bibnamefont
  {Gonz\'{a}lez-Tudela}}\ and\ \bibinfo {author} {\bibfnamefont {J.~I.}\
  \bibnamefont {Cirac}},\ }\bibfield  {title} {\bibinfo {title} {Quantum
  {E}mitters in {T}wo-{D}imensional {S}tructured {R}eservoirs in the
  {N}onperturbative {R}egime},\ }\href
  {https://journals.aps.org/prl/abstract/10.1103/PhysRevLett.119.143602}
  {\bibfield  {journal} {\bibinfo  {journal} {Phys. Rev. Lett.}\ }\textbf
  {\bibinfo {volume} {119}},\ \bibinfo {pages} {143602} (\bibinfo {year}
  {2017}{\natexlab{a}})}\BibitemShut {NoStop}%
\bibitem [{\citenamefont {Gonz\'{a}lez-Tudela}\ and\ \citenamefont
  {Cirac}(2017{\natexlab{b}})}]{AGT2017pra}%
  \BibitemOpen
  \bibfield  {author} {\bibinfo {author} {\bibfnamefont {A.}~\bibnamefont
  {Gonz\'{a}lez-Tudela}}\ and\ \bibinfo {author} {\bibfnamefont {J.~I.}\
  \bibnamefont {Cirac}},\ }\bibfield  {title} {\bibinfo {title} {Markovian and
  non-{M}arkovian dynamics of quantum emitters coupled to two-dimensional
  structured reservoirs},\ }\href
  {https://journals.aps.org/pra/abstract/10.1103/PhysRevA.96.043811} {\bibfield
   {journal} {\bibinfo  {journal} {Phys. Rev. A}\ }\textbf {\bibinfo {volume}
  {96}},\ \bibinfo {pages} {043811} (\bibinfo {year}
  {2017}{\natexlab{b}})}\BibitemShut {NoStop}%
\bibitem [{\citenamefont {Zhao}\ and\ \citenamefont {Wang}(2020)}]{ZhaoWbound}%
  \BibitemOpen
  \bibfield  {author} {\bibinfo {author} {\bibfnamefont {W.}~\bibnamefont
  {Zhao}}\ and\ \bibinfo {author} {\bibfnamefont {Z.}~\bibnamefont {Wang}},\
  }\bibfield  {title} {\bibinfo {title} {Single-photon scattering and bound
  states in an atom-waveguide system with two or multiple coupling points},\
  }\href {https://journals.aps.org/pra/abstract/10.1103/PhysRevA.101.053855}
  {\bibfield  {journal} {\bibinfo  {journal} {Phys. Rev. A}\ }\textbf {\bibinfo
  {volume} {101}},\ \bibinfo {pages} {053855} (\bibinfo {year}
  {2020})}\BibitemShut {NoStop}%
\bibitem [{\citenamefont {Wang}\ and\ \citenamefont {Wan}(2022)}]{DualityPRB}%
  \BibitemOpen
  \bibfield  {author} {\bibinfo {author} {\bibfnamefont {S.-X.}\ \bibnamefont
  {Wang}}\ and\ \bibinfo {author} {\bibfnamefont {S.}~\bibnamefont {Wan}},\
  }\bibfield  {title} {\bibinfo {title} {Duality between the generalized
  non-{H}ermitian {H}atano-{N}elson model in flat space and a {H}ermitian
  system in curved space},\ }\href
  {https://journals.aps.org/prb/abstract/10.1103/PhysRevB.106.075112}
  {\bibfield  {journal} {\bibinfo  {journal} {Phys. Rev. B}\ }\textbf {\bibinfo
  {volume} {106}},\ \bibinfo {pages} {075112} (\bibinfo {year}
  {2022})}\BibitemShut {NoStop}%
\bibitem [{\citenamefont {Koll\'{a}r}\ \emph {et~al.}(2019)\citenamefont
  {Koll\'{a}r}, \citenamefont {Fitzpatrick},\ and\ \citenamefont
  {Houck}}]{Ccurve1}%
  \BibitemOpen
  \bibfield  {author} {\bibinfo {author} {\bibfnamefont {A.~J.}\ \bibnamefont
  {Koll\'{a}r}}, \bibinfo {author} {\bibfnamefont {M.}~\bibnamefont
  {Fitzpatrick}},\ and\ \bibinfo {author} {\bibfnamefont {A.~A.}\ \bibnamefont
  {Houck}},\ }\bibfield  {title} {\bibinfo {title} {Hyperbolic lattices in
  circuit quantum electrodynamics},\ }\href
  {https://www.nature.com/articles/s41586-019-1348-3} {\bibfield  {journal}
  {\bibinfo  {journal} {Nature (London)}\ }\textbf {\bibinfo {volume} {571}},\
  \bibinfo {pages} {45} (\bibinfo {year} {2019})}\BibitemShut {NoStop}%
\bibitem [{\citenamefont {Bienias}\ \emph {et~al.}(2022)\citenamefont
  {Bienias}, \citenamefont {Boettcher}, \citenamefont {Belyansky},
  \citenamefont {Koll\'{a}r},\ and\ \citenamefont {Gorshkov}}]{Ccurve2}%
  \BibitemOpen
  \bibfield  {author} {\bibinfo {author} {\bibfnamefont {P.}~\bibnamefont
  {Bienias}}, \bibinfo {author} {\bibfnamefont {I.}~\bibnamefont {Boettcher}},
  \bibinfo {author} {\bibfnamefont {R.}~\bibnamefont {Belyansky}}, \bibinfo
  {author} {\bibfnamefont {A.~J.}\ \bibnamefont {Koll\'{a}r}},\ and\ \bibinfo
  {author} {\bibfnamefont {A.~V.}\ \bibnamefont {Gorshkov}},\ }\bibfield
  {title} {\bibinfo {title} {Circuit {Q}uantum {E}lectrodynamics in
  {H}yperbolic {S}pace: {F}rom {P}hoton {B}ound {S}tates to {F}rustrated {S}pin
  {M}odels},\ }\href
  {https://journals.aps.org/prl/abstract/10.1103/PhysRevLett.128.013601}
  {\bibfield  {journal} {\bibinfo  {journal} {Phys. Rev. Lett.}\ }\textbf
  {\bibinfo {volume} {128}},\ \bibinfo {pages} {013601} (\bibinfo {year}
  {2022})}\BibitemShut {NoStop}%
\bibitem [{\citenamefont {Gonz\'{a}lez-Tudela}\ \emph
  {et~al.}(2019)\citenamefont {Gonz\'{a}lez-Tudela}, \citenamefont
  {Mu\~{n}oz},\ and\ \citenamefont {Cirac}}]{AGT2D1}%
  \BibitemOpen
  \bibfield  {author} {\bibinfo {author} {\bibfnamefont {A.}~\bibnamefont
  {Gonz\'{a}lez-Tudela}}, \bibinfo {author} {\bibfnamefont {C.~S.}\
  \bibnamefont {Mu\~{n}oz}},\ and\ \bibinfo {author} {\bibfnamefont {J.~I.}\
  \bibnamefont {Cirac}},\ }\bibfield  {title} {\bibinfo {title} {Engineering
  and {H}arnessing {G}iant {A}toms in {H}igh-{D}imensional {B}aths: {A}
  {P}roposal for {I}mplementation with {C}old {A}toms},\ }\href
  {https://journals.aps.org/prl/abstract/10.1103/PhysRevLett.122.203603}
  {\bibfield  {journal} {\bibinfo  {journal} {Phys. Rev. Lett.}\ }\textbf
  {\bibinfo {volume} {122}},\ \bibinfo {pages} {203603} (\bibinfo {year}
  {2019})}\BibitemShut {NoStop}%
\bibitem [{\citenamefont {Vega}\ \emph {et~al.}(2023)\citenamefont {Vega},
  \citenamefont {Porras},\ and\ \citenamefont {Gonz\'{a}lez-Tudela}}]{AGT2D2}%
  \BibitemOpen
  \bibfield  {author} {\bibinfo {author} {\bibfnamefont {C.}~\bibnamefont
  {Vega}}, \bibinfo {author} {\bibfnamefont {D.}~\bibnamefont {Porras}},\ and\
  \bibinfo {author} {\bibfnamefont {A.}~\bibnamefont {Gonz\'{a}lez-Tudela}},\
  }\bibfield  {title} {\bibinfo {title} {Topological multimode waveguide
  {QED}},\ }\href
  {https://journals.aps.org/prresearch/abstract/10.1103/PhysRevResearch.5.023031}
  {\bibfield  {journal} {\bibinfo  {journal} {Phys. Rev. Res.}\ }\textbf
  {\bibinfo {volume} {5}},\ \bibinfo {pages} {023031} (\bibinfo {year}
  {2023})}\BibitemShut {NoStop}%
\bibitem [{\citenamefont {Ingelsten}\ \emph {et~al.}()\citenamefont
  {Ingelsten}, \citenamefont {Kockum},\ and\ \citenamefont {Soro}}]{AFK2D}%
  \BibitemOpen
  \bibfield  {author} {\bibinfo {author} {\bibfnamefont {E.}~\bibnamefont
  {Ingelsten}}, \bibinfo {author} {\bibfnamefont {A.~F.}\ \bibnamefont
  {Kockum}},\ and\ \bibinfo {author} {\bibfnamefont {A.}~\bibnamefont {Soro}},\
  }\bibfield  {title} {\bibinfo {title} {Giant atoms in a two-dimensional
  structured environment},\ }\href@noop {} {\bibinfo  {journal} {in
  preparation}\ }\BibitemShut {NoStop}%
\bibitem [{\citenamefont {Feng}\ \emph
  {et~al.}(2017{\natexlab{b}})\citenamefont {Feng}, \citenamefont
  {El-Ganainy},\ and\ \citenamefont {Ge}}]{FengNP2017}%
  \BibitemOpen
\bibfield  {journal} {  }\bibfield  {author} {\bibinfo {author} {\bibfnamefont
  {L.}~\bibnamefont {Feng}}, \bibinfo {author} {\bibfnamefont {R.}~\bibnamefont
  {El-Ganainy}},\ and\ \bibinfo {author} {\bibfnamefont {L.}~\bibnamefont
  {Ge}},\ }\bibfield  {title} {\bibinfo {title} {Non-{H}ermitian photonics
  based on parity–time symmetry},\ }\href
  {https://www.nature.com/articles/s41566-017-0031-1} {\bibfield  {journal}
  {\bibinfo  {journal} {Nat. Photon.}\ }\textbf {\bibinfo {volume} {11}},\
  \bibinfo {pages} {752} (\bibinfo {year} {2017}{\natexlab{b}})}\BibitemShut
  {NoStop}%
\bibitem [{\citenamefont {Bekenstein}\ \emph {et~al.}(2017)\citenamefont
  {Bekenstein}, \citenamefont {Kabessa}, \citenamefont {Sharabi}, \citenamefont
  {Tal}, \citenamefont {Engheta}, \citenamefont {Eisenstein}, \citenamefont
  {Agranat},\ and\ \citenamefont {Segev}}]{Ccurve3}%
  \BibitemOpen
  \bibfield  {author} {\bibinfo {author} {\bibfnamefont {R.}~\bibnamefont
  {Bekenstein}}, \bibinfo {author} {\bibfnamefont {Y.}~\bibnamefont {Kabessa}},
  \bibinfo {author} {\bibfnamefont {Y.}~\bibnamefont {Sharabi}}, \bibinfo
  {author} {\bibfnamefont {O.}~\bibnamefont {Tal}}, \bibinfo {author}
  {\bibfnamefont {N.}~\bibnamefont {Engheta}}, \bibinfo {author} {\bibfnamefont
  {G.}~\bibnamefont {Eisenstein}}, \bibinfo {author} {\bibfnamefont {A.~J.}\
  \bibnamefont {Agranat}},\ and\ \bibinfo {author} {\bibfnamefont
  {M.}~\bibnamefont {Segev}},\ }\bibfield  {title} {\bibinfo {title} {Control
  of light by curved space in nanophotonic structures},\ }\href
  {https://www.nature.com/articles/s41566-017-0008-0} {\bibfield  {journal}
  {\bibinfo  {journal} {Nat. Photon.}\ }\textbf {\bibinfo {volume} {11}},\
  \bibinfo {pages} {664} (\bibinfo {year} {2017})}\BibitemShut {NoStop}%
\bibitem [{\citenamefont {Schine}\ \emph {et~al.}(2019)\citenamefont {Schine},
  \citenamefont {Chalupnik}, \citenamefont {Can}, \citenamefont {Gromov},\ and\
  \citenamefont {Simon}}]{Ccurve4}%
  \BibitemOpen
  \bibfield  {author} {\bibinfo {author} {\bibfnamefont {N.}~\bibnamefont
  {Schine}}, \bibinfo {author} {\bibfnamefont {M.}~\bibnamefont {Chalupnik}},
  \bibinfo {author} {\bibfnamefont {T.}~\bibnamefont {Can}}, \bibinfo {author}
  {\bibfnamefont {A.}~\bibnamefont {Gromov}},\ and\ \bibinfo {author}
  {\bibfnamefont {J.}~\bibnamefont {Simon}},\ }\bibfield  {title} {\bibinfo
  {title} {Electromagnetic and gravitational responses of photonic {L}andau
  levels},\ }\href {https://www.nature.com/articles/s41586-018-0817-4}
  {\bibfield  {journal} {\bibinfo  {journal} {Nature (London)}\ }\textbf
  {\bibinfo {volume} {565}},\ \bibinfo {pages} {173} (\bibinfo {year}
  {2019})}\BibitemShut {NoStop}%
\bibitem [{\citenamefont {Zhang}\ \emph {et~al.}(2021)\citenamefont {Zhang},
  \citenamefont {Lv}, \citenamefont {Yan},\ and\ \citenamefont
  {Zhou}}]{Bulletin}%
  \BibitemOpen
  \bibfield  {author} {\bibinfo {author} {\bibfnamefont {R.}~\bibnamefont
  {Zhang}}, \bibinfo {author} {\bibfnamefont {C.}~\bibnamefont {Lv}}, \bibinfo
  {author} {\bibfnamefont {Y.}~\bibnamefont {Yan}},\ and\ \bibinfo {author}
  {\bibfnamefont {Q.}~\bibnamefont {Zhou}},\ }\bibfield  {title} {\bibinfo
  {title} {Efimov-like states and quantum funneling effects on synthetic
  hyperbolic surfaces},\ }\href
  {https://www.sciengine.com/SB/doi/10.1016/j.scib.2021.06.017} {\bibfield
  {journal} {\bibinfo  {journal} {Sci. Bull.}\ }\textbf {\bibinfo {volume}
  {66}},\ \bibinfo {pages} {1967} (\bibinfo {year} {2021})}\BibitemShut
  {NoStop}%
\end{thebibliography}%

\clearpage

\onecolumngrid

\begin{center}
{\large \textbf{Supplemental Material for ``Giant Emitters in a Structured Bath with Non-Hermitian Skin Effect''}}

\vspace{8mm}

Lei Du$^{1,2}$, Lingzhen Guo$^{3}$, Yan Zhang$^{1}$, and Anton Frisk Kockum$^{2}$
\end{center}

\begin{minipage}[]{16cm}
\small{\it

\centering $^{1}$ Center for Quantum Sciences and School of Physics, \\
\centering Northeast Normal University, Changchun 130024, China  \\
\centering $^{2}$ Department of Microtechnology and Nanoscience, \\
\centering Chalmers University of Technology, 412 96 Gothenburg, Sweden  \\
\centering $^{3}$ Center for Joint Quantum Studies and Department of Physics, School of Science, \\
\centering Tianjin University, Tianjin 300072, China \\
}

\end{minipage}

\vspace{8mm}

\setcounter{figure}{0}
\renewcommand{\thefigure}{S\arabic{figure}}
\setcounter{equation}{0}
\renewcommand{\theequation}{S\arabic{equation}}

\section{I.\quad Self-energy of a giant emitter coupled to a Hatano-Nelson model}

The dynamics of the single giant emitter $b$ can be clearly illustrated using the resolvent method~\cite{resolvent}. The resolvent of the whole system (i.e., emitter$+$bath) is
\begin{equation}
\mathbb{G}(z)=\frac{1}{z-H_{0}-H_{\text{int}}},
\label{resolvent}
\end{equation}
where $H_{0}$ is the free Hamiltonian of the emitter and the bare Hatano-Nelson (HN) model and $H_{\text{int}}$ describes the emitter-lattice interaction. For a projector $P$ of the eigenstates of $H_{0}$ and its complementary $Q=1-P$, the resolvent satisfies
\begin{equation}
P\mathbb{G}(z)P=\frac{P}{z-PH_{0}P-P\Sigma(z)P},
\label{project}
\end{equation}
where
\begin{equation}
\begin{split}
\Sigma(z)&=H_{\text{int}}+H_{\text{int}}\frac{Q}{z-QH_{0}Q-QH_{\text{int}}Q}H_{\text{int}}\\
&\simeq H_{\text{int}}+H_{\text{int}}\frac{Q}{z-H_{0}}H_{\text{int}}
\end{split}
\label{levelshift}
\end{equation}
 is the so-called level-shift operator. Note that the second-order perturbation expansion in Eq.~(\ref{levelshift}) is justified if $H_{\text{int}}$ is small compared to $H_{0}$. By using the projector $P=|b\rangle\langle b|$ (projecting onto the state of emitter $b$) and its complementary $Q=1-P=\sum_{q}|q\rangle\langle q|$ (projecting onto the states of the bath), the self-energy of $b$ can be given by $\Sigma_{b}(z)=\langle b|\Sigma(z)|b\rangle$.

If both $t_{R}$ and $t_{L}$ are positive (i.e., the HN model is in the \emph{convectively unstable regime}~\cite{LonghiQD}), the Hamiltonian of the HN model can be rewritten as~\cite{Longhi2015prb}
\begin{equation}
H_{\text{HN}}=\sum_{n}\omega_{0}a_{n}^{\dag}a_{n}+\sum_{n}\mleft(\sqrt{t_{R}t_{L}}e^{\text{ln}\beta}a_{n+1}^{\dag}a_{n}+\sqrt{t_{R}t_{L}}e^{-\text{ln}\beta}a_{n}^{\dag}a_{n+1}\mright).
\label{SMHNH}
\end{equation}
Its eigenvalues are given by
\begin{equation}
E_{q}=\omega_{0}+2\sqrt{t_{R}t_{L}}\cos{q},
\label{SMHNE}
\end{equation}
where $q=k-i\ln(\beta)=k-i\ln(\sqrt{t_{R}/t_{L}})$ is a \emph{complex} wave number and $\omega_{0}$ is the resonance frequency of each lattice site as well as emitter $b$ (we assume that the emitter is resonant with the lattice band center). This implies that the HN model can be viewed as a one-dimensional pseudo-Hermitian lattice subject to an \emph{imaginary} gauge field. In view of this, by performing the transformation
\begin{equation}
a_{q}=\frac{1}{\sqrt{2\pi}}\sum_{n}a_{n}e^{-iqn},
\label{transform}
\end{equation}
the Hamiltonian in Eq.~(2) in the main text can be rewritten as
\begin{eqnarray}
 \tilde{H}(t)&=&H_{0}+H_{\text{int}}, 
 \label{SMkH}
\\
 H_{0}&=&\sum_{q}\Delta_{k}a_{q}^{\dag}a_{q},
 \label{SMH0}
\\
 H_{\text{int}}&=&\sum_{q}\left[\left(G_{N}+G_{N'}e^{-iDq}\right)e^{-iNq}ba_{q}^{\dag}+\left(G_{N}+G_{N'}e^{iDq}\right)e^{iNq}b^{\dag}a_{q}\right],
 \label{SMHint}
\end{eqnarray}
with $G_{N}=g_{N}/\sqrt{2\pi}$, $G_{N'}=g_{N'}/\sqrt{2\pi}$, $D=N'-N$, and $\Delta_{k}=2\sqrt{t_{R}t_{L}}\cos{k}$. Here $a_{q}$ is the $q$-space annihilation operator of the lattice field, satisfying $[a_{q},a_{q'}^{\dag}]=\delta(q-q')$ and $[a_{q},a_{q'}]=[a_{q}^{\dag},a_{q'}^{\dag}]=0$. It is worth noting that $\Delta_{k}$ is real and in general not a function of $\beta$.

Substituting Eqs.~(\ref{SMH0}) and (\ref{SMHint}) into Eq.~(\ref{levelshift}), the self-energy of $b$ is obtained as
\begin{equation}
\begin{split}
\Sigma_{b}(z)&=\sum_{q}\frac{\mleft(G_{N}+G_{N'}e^{-iDq}\mright)\mleft(G_{N}+G_{N'}e^{iDq}\mright)}{z-\Delta_{k}}\\
&=\frac{1}{2\pi}\int dq\frac{\mleft(G_{N}+G_{N'}e^{-iDq}\mright)\mleft(G_{N}+G_{N'}e^{iDq}\mright)}{z-2\sqrt{t_{R}t_{L}}\cos{k}}\\
&=\frac{1}{2\pi}\int dk\frac{G_{N}^{2}+G_{N'}^{2}+G_{N}G_{N'}e^{iDk}\mleft(\beta^{D}+\beta^{-D}\mright)}{z-2\sqrt{t_{R}t_{L}}\cos{k}},
\end{split}
\label{self1}
\end{equation}
where in the last step we have changed the integral variable as $\int dq\rightarrow\int dk$ and replaced $\text{exp}(-iDk)$ by $\text{exp}(iDk)$ (only even functions contribute to the integral). By setting $y=\text{exp}(ik)$ such that $2\cos{k}=y+y^{-1}$ and $\int_{-\pi}^{+\pi}dk=-i\oint y^{-1}dy$, we have
\begin{equation}
\Sigma_{b}(z)=\frac{-i}{2\pi}\oint dy\frac{G_{N}^{2}+G_{N'}^{2}+G_{N}G_{N'}y^{D}\mleft(\beta^{D}+\beta^{-D}\mright)}{zy-\sqrt{t_{R}t_{L}}(y^{2}+1)}.
\label{self2}
\end{equation}
By using the residue theorem, we finally arrive at
\begin{equation}
\Sigma_{b}(z)=\mp\frac{1}{\sqrt{z^{2}-4t_{R}t_{L}}}\mleft[G_{N}^{2}+G_{N'}^{2}+G_{N}G_{N'}y_{\pm}^{D}\mleft(\beta^{D}+\beta^{-D}\mright)\mright]
\label{self3}
\end{equation}
with $y_{\pm}=[z\pm\sqrt{z^{2}-4t_{R}t_{L}})]/(2\sqrt{t_{R}t_{L}})$. In the case of weak emitter-bath couplings (i.e., under the Weisskopf-Wigner approximation), the dynamics of the emitter are well captured by the self-energy close to the real axis~\cite{AGT2017prl,AGT2017pra}, i.e., $\Sigma_{b}(\Delta+i0^{+})$ with $\Delta$ the frequency detuning between the emitter and the middle of the energy band. If $b$ is exactly resonant with the band center (i.e., $\Delta=0$), one has $y_{\pm}=\pm i$ and Eq.~(\ref{self3}) can be simplified to
\begin{equation}
\Sigma_{b}(\Delta=0)\approx\frac{\pm i}{2\sqrt{t_{R}t_{L}}}\mleft[G_{N}^{2}+G_{N'}^{2}+(\pm i)^{D}G_{N}G_{N'}\mleft(\beta^{D}+\beta^{-D}\mright)\mright].
\label{simself}
\end{equation}
Clearly, Eq.~(\ref{simself}) can correctly predict the dynamics in Fig.~1 in the main text if $G_{N'}=G_{N}\beta^{-D}$.

\section{II.\quad Nonreciprocal decoherence-free interaction in the braided double-giant-emitter model}

In this section, we aim to calculate the effective interaction between the two braided giant emitters $b$ and $c$ considered in the main text. By using the transformation in Eq.~(\ref{transform}), the Hamiltonian in Eq.~(4) in the main text becomes $\tilde{H}'=H_{0}+H_{\text{int}}'$, where $H_{0}$ is given in Eq.~(\ref{SMH0}) and
\begin{equation}
\begin{split}
H_{\text{int}}'&=\sum_{q}\mleft[\mleft(G_{N}+G_{N+2}e^{-2iq}\mright)e^{-iNq}ba_{q}^{\dag}+\mleft(G_{N+1}e^{-iq}+G_{N+3}e^{-3iq}\mright)e^{-iNq}ca_{q}^{\dag}\mright.\\
 &\mleft.\quad\,+\mleft(G_{N}+G_{N+2}e^{2iq}\mright)e^{iNq}b^{\dag}a_{q}+\mleft(G_{N+1}e^{iq}+G_{N+3}e^{3iq}\mright)e^{iNq}c^{\dag}a_{q}\mright]
 \end{split}
 \label{SMdimerH}
\end{equation}
with $G_{N+j}=g_{N+j}/\sqrt{2\pi}$ ($j=0,1,2,3$). Note that in Eq.~(\ref{SMdimerH}) the last two terms are not the Hermitian conjugates of the first two ones since $q$ is complex.

Once again, by substituting Eqs.~(\ref{SMH0}) and (\ref{SMdimerH}) into Eq.~(\ref{resolvent}), the resolvent of the two-giant-emitters model can be transformed by the projector $P=|b\rangle\langle b|+|c\rangle\langle c|$ (projecting onto the states of the emitters $b$ and $c$) into a $2\times2$ matrix form, i.e.,
\begin{equation}
P\mathbb{G}(z)P=
\begin{pmatrix}
G_{bb} & G_{bc} \\
G_{cb} & G_{cc}
\end{pmatrix}
=
\begin{pmatrix}
z-\Sigma_{b} & -\Sigma_{bc} \\
-\Sigma_{cb} & z-\Sigma_{cc}
\end{pmatrix},
\label{project2}
\end{equation}
where $\sum_{\alpha}=\langle \alpha|\Sigma(z)|\alpha\rangle$, $\sum_{\alpha\alpha'}=\langle \alpha|\Sigma(z)|\alpha'\rangle$ ($\alpha,\,\alpha'=b,\,c$), and $Q=1-P=\sum_{q}|q\rangle\langle q|$. If $G_{N}=G_{N+1}$ and $G_{N+2}=G_{N+3}=G_{N}/\beta^{2}$ as assumed in the main text, the self-energy of $b$ can be obtained as
\begin{equation}
\Sigma_{b}(z)=\mp\frac{G_{N}^{2}}{\sqrt{z^{2}-4t_{R}t_{L}}}\mleft[1+y_{\pm}^{2}+\frac{1}{\beta^{4}}\mleft(1+y_{\pm}^{2}\mright)\mright],
\label{simself2}
\end{equation}
which is identical with that of the single-emitter case [cf.~Eq.~(\ref{self3}) for $G_{N'}=G_{N}/\beta^{D}$ and $D=2$]. Since the emitters $b$ and $c$ are identical, one can readily verify that $\Sigma_{c}(z)=\Sigma_{b}(z)$. Clearly, the individual relaxations of $b$ and $c$, which are described by the imaginary parts of $\Sigma_{b}(\Delta+i0^{+})$ and $\Sigma_{c}(\Delta+i0^{+})$, are inhibited if both $b$ and $c$ are resonant with the band center of the HN model.

Moreover, the interaction parts $\Sigma_{bc}(z)$ and $\Sigma_{cb}(z)$, i.e., the off-diagonal elements of Eq.~(\ref{project2}), can be obtained as
\begin{equation}
\begin{split}
\Sigma_{bc}(z)&=G_{N}^{2}\sum_{q}\frac{\mleft(e^{-iq}+\frac{1}{\beta^{2}}e^{-3iq}\mright)\mleft(1+\frac{1}{\beta^{2}}e^{2iq}\mright)}{z-\Delta_{k}}\\
&=\frac{G_{N}^{2}}{2\pi}\int dk\frac{\frac{1}{\beta}\mleft(e^{ik}+e^{-ik}\mright)+\frac{1}{\beta^{5}}\mleft(e^{-3ik}+e^{-ik}\mright)}{z-2\sqrt{t_{R}t_{L}}\cos{k}}\\
&=\frac{G_{N}^{2}}{2\pi}\int dk\frac{\frac{2}{\beta}e^{ik}+\frac{1}{\beta^{5}}\mleft(e^{3ik}+e^{ik}\mright)}{z-2\sqrt{t_{R}t_{L}}\cos{k}}\\
&=\mp\frac{G_{N}^{2}}{\sqrt{z^{2}-4t_{R}t_{L}}}\mleft[\frac{2}{\beta}y_{\pm}+\frac{1}{\beta^{5}}\mleft(y_{\pm}+y_{\pm}^{3}\mright)\mright]
\end{split}
\label{cross1}
\end{equation}
and
\begin{equation}
\begin{split}
\Sigma_{cb}(z)&=G_{N}^{2}\sum_{q}\frac{\mleft(e^{iq}+\frac{1}{\beta^{2}}e^{3iq}\mright)\mleft(1+\frac{1}{\beta^{2}}e^{-2iq}\mright)}{z-\Delta_{k}}\\
&=\frac{G_{N}^{2}}{2\pi}\int dk\frac{\beta\mleft(e^{ik}+e^{3ik}\mright)+\frac{1}{\beta^{3}}\mleft(e^{ik}+e^{-ik}\mright)}{z-2\sqrt{t_{R}t_{L}}\cos{k}}\\
&=\frac{G_{N}^{2}}{2\pi}\int dk\frac{\beta\mleft(e^{ik}+e^{3ik}\mright)+\frac{2}{\beta^{3}}e^{ik}}{z-2\sqrt{t_{R}t_{L}}\cos{k}}\\
&=\mp\frac{G_{N}^{2}}{\sqrt{z^{2}-4t_{R}t_{L}}}\mleft[\beta\mleft(y_{\pm}+y_{\pm}^{3}\mright)+\frac{2}{\beta^{3}}y_{\pm}\mright].
\end{split}
\label{cross2}
\end{equation}
When $b$ and $c$ are resonant with the band center of the HN model and under the Weisskopf-Wigner approximation, we have
\begin{eqnarray}
\Sigma_{bc}(z)\rightarrow\Sigma_{bc}(0+i0^{+})&\approx&\frac{-G_{N}^{2}}{\beta\sqrt{t_{R}t_{L}}},
\label{simcross1}
\\
\Sigma_{cb}(z)\rightarrow\Sigma_{cb}(0+i0^{+})&\approx&\frac{-G_{N}^{2}}{\beta^{3}\sqrt{t_{R}t_{L}}}, 
\label{simcross2}
\end{eqnarray}
which are purely real (keeping in mind that $\{t_{R},\,t_{L}\}>0$ has been assumed). Since the real and imaginary parts of $\Sigma_{bc(cb)}(z)$ describe the effective coherent coupling from $c$ to $b$ (from $b$ to $c$) and the collective relaxation of the two emitters, respectively, we conclude from Eqs.~(\ref{simcross1}) and (\ref{simcross2}) that there is a nonreciprocal decoherence-free interaction (with a weight difference $\beta^{-2}$) between $b$ and $c$.

With a similar procedure, we can also calculate the interaction parts $\Sigma_{bc}(z)$ and $\Sigma_{cb}(z)$ in a more general case with \emph{arbitrary} coupling separations [beyond the nearest-neighbor coupling points in Eq.~(\ref{SMdimerH})]. For example, we can immediately obtain
\begin{eqnarray}
\Sigma_{bc}(0+i0^{+})&\approx&\mp\frac{G_{N}^{2}}{2i\sqrt{t_{R}t_{L}}}\frac{1}{\beta^{D'}}\mleft[\mleft(\pm i\mright)^{D'}-\mleft(\pm i\mright)^{-D'}\mright],
\label{absep1}
\\
\Sigma_{cb}(0+i0^{+})&\approx&\mp\frac{G_{N}^{2}}{2i\sqrt{t_{R}t_{L}}}\frac{\beta^{D'}}{\beta^{2D''}}\mleft[\mleft(\pm i\mright)^{D'}-\mleft(\pm i\mright)^{-D'}\mright],
\label{absep2}
\end{eqnarray}
where $D'$ is the separation between the left coupling points of the two giant emitters [$D'=1$ in Eqs.~(\ref{simcross1}) and (\ref{simcross2})] and $D''$ is the separation between the two coupling points of each giant emitter [$D''=2$ in Eqs.~(\ref{simcross1}) and (\ref{simcross2})]. For simplicity, we have assumed identical separation $D''$ for both emitters. It is clear from Eqs.~(\ref{absep1}) and (\ref{absep2}) that if $(\pm i)^{D'}\neq 1$, there is always a nonreciprocal decoherence-free interaction which is manifested by the strength difference $\beta^{2(D'-D'')}$. However, the coupling strengths decay with the increase of $D'$ and $D''$ and the decay rate is determined by the non-Hermiticity $\gamma$ (i.e., the value of $\beta$).

\section{III.\quad Hidden bound states}

\begin{figure}
\centering
\includegraphics[width=0.6\linewidth]{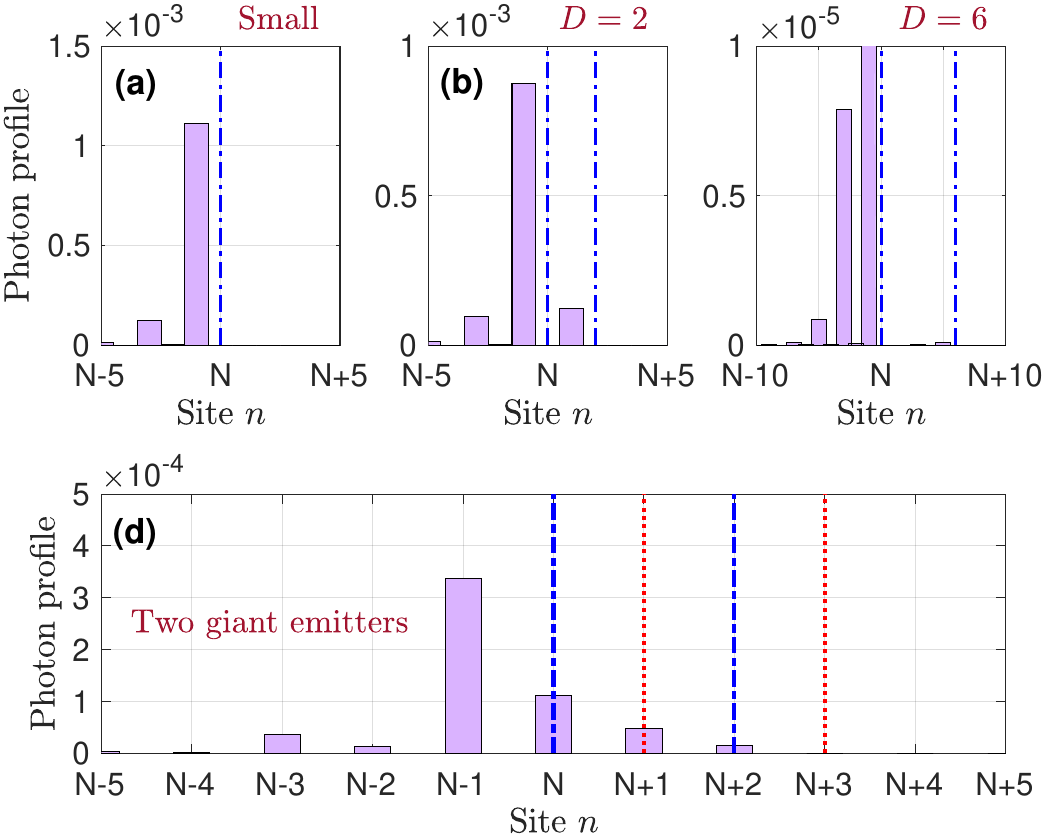}
\caption{Photon profiles of the hidden bound state for different models: (a) a single small emitter, (b, c) a single giant emitter with different values of coupling separation, and (d) two braided giant emitters as considered in Fig.~2 in the main text. The vertical lines indicate the coupling points of the giant emitters, with different line styles (and colors) corresponding to different emitters. Other parameters are $g_{N}=1$, $\nu=20$, $\gamma=10$, and $M=1000$.}
\label{FhiddenNew}
\end{figure}

It has been shown that there are two essentially different bound states that can arise when considering a small emitter coupled to an HN model~\cite{GongHN}. If the frequency of the emitter falls within the energy band of the HN model (within the energy loop in the complex plane), there is always a \emph{hidden bound state} localized around the emitter-lattice coupling point, whose energy is exactly pinned at the frequency of the emitter. This bound state shows a \emph{chiral} photonic profile which decays exponentially in the direction of the weaker tunneling rate, as shown in \figpanel{FhiddenNew}{a}. On the other hand, conventional bound states can be formed if the frequency of the emitter falls outside the energy band. In contrast to the Hermitian case, the conventional bound states here appear only if the emitter-lattice coupling strength is large enough. If several small emitters (which can have different frequencies) are coupled to a common HN model, the photonic profile of the hidden bound state is a superposition of those arising from each atom, as if the emitters do not influence each other~\cite{GongHN2}.

In this section, we briefly discuss the hidden bound states induced by a giant emitter coupled to an HN model. For a single giant emitter, as shown in \figpanel{FhiddenNew}{b}, the photonic profile of the resulting hidden bound state is also a superposition of those arising from each coupling point. However, the components induced by the two coupling points are quite imbalanced due to the unequal coupling strengths. With the matching condition $g_{N'}/g_{N}=\beta^{-D}$ fulfilled, the component induced by the right coupling point diminishes rapidly as the coupling separation $D$ increases, as shown in \figpanel{FhiddenNew}{c}. 


In fact, the hidden bound states can provide an intuitive but qualitative picture to understand the nonreciprocal decoherence-free interaction in Fig.~2 in the main text (whose nonreciprocity is opposite to that of the HN model). As discussed above, each giant emitter can produce a hidden bound state with a highly chiral photonic profile with respect to its two coupling points. If an emitter $A$ is located in the range of the bound state induced by another emitter $B$, the intensity of the bound state at the coupling point of $A$ determines the effective coupling strength from $B$ to $A$ through this coupling path. In the case of Fig.~2 in the main text, the intensities of the hidden bound state (i.e., induced by $b$) at the coupling points of $c$ are smaller than the intensities of the hidden bound state (i.e., induced by $c$) at the coupling points of $b$, as shown in \figpanel{FhiddenNew}{d}. Keeping in mind that this is almost the only path that mediates the interaction between the two emitters (both of them nearly do not decay into the lattice), they thus exhibit nonreciprocal interaction with opposite direction preference with respect to the HN model.

\section{IV.\quad Dynamics of a small emitter in the convectively and absolutely unstable regimes}

\begin{figure}
\centering
\includegraphics[width = 0.6\linewidth]{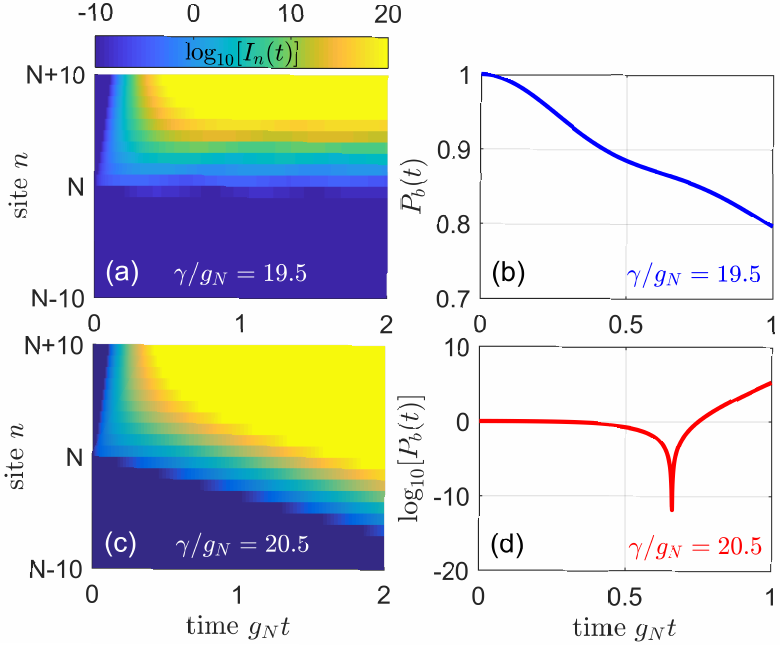}
\caption{(a, c) Time evolution of the field intensities $I_{n}(t)$ of the lattice sites for (a) $\gamma/g_{N}=19.5$ and (c) $\gamma/g_{N}=20.5$. (b, d) Time evolution of the mean particle number $P_{b}(t)$ of the small emitter for (b) $\gamma/g_{N}=19.5$ and (d) $\gamma/g_{N}=20.5$. We assume that the small emitter is coupled to the $N$th site of the HN model. Other parameters are $g_{N}=1$, $\nu=20$, and $M=1000$.}
\label{Fsmall}
\end{figure}

As discussed in Sec.~I, an HN model is equivalent to a pseudo-Hermitian lattice subject to an imaginary gauge field when $\{t_{R},\,t_{L}\}>0$ (i.e., in the convectively unstable regime). The imaginary gauge field, which is manifested by the complex wave number $q=k-i\ln(\beta)$, can be eliminated by performing the transformation
\begin{equation}
\tilde{a}_{n}=a_{n}\exp\mleft[-n\ln(\beta)\mright].
\label{eliminate}
\end{equation}
In this regime, the relaxation dynamics of a small emitter resembles the Hermitian case. As shown in \figpanel{Fsmall}{a}, although the field is amplified when traveling along the direction of the larger tunneling rate, it can be transferred away rapidly with almost no remnant at each passing lattice site in the long-time limit [i.e., for any given lattice site $a_{n}$, its field intensity $I_{n}(t)=|u_{a,n}(t)|^{2}\rightarrow0$ for $t\rightarrow+\infty$]. This implies that the instability is \emph{transient} in this case~\cite{LonghiQD}. Therefore, the emitter shows an irreversible complete decay if its transition frequency lies within the energy band of the HN model, as shown in \figpanel{Fsmall}{b}.

In the absolutely unstable regime (i.e., $t_{R}>0$ and $t_{L}<0$), however, the HN model cannot be mapped to a pseudo-Hermitian one. In this case, the decay-dynamics problem becomes purely non-Hermitian. As shown in \figpanel{Fsmall}{c}, the excitation is strongly amplified after entering the HN model, similar to the convectively unstable case. However, at the coupled site $n=N$ there is always a considerable remnant suffering permanent amplification [i.e., $I_{N}(t)\rightarrow+\infty$ for $t\rightarrow+\infty$]. In other words, the local amplification effect surpasses the effective transport capability of the HN model so that the emitter shows a secular energy growth, as shown in \figpanel{Fsmall}{d}.

\section{V.\quad Mapping to curved spaces}

\begin{figure}
\centering
\includegraphics[width=0.6\linewidth]{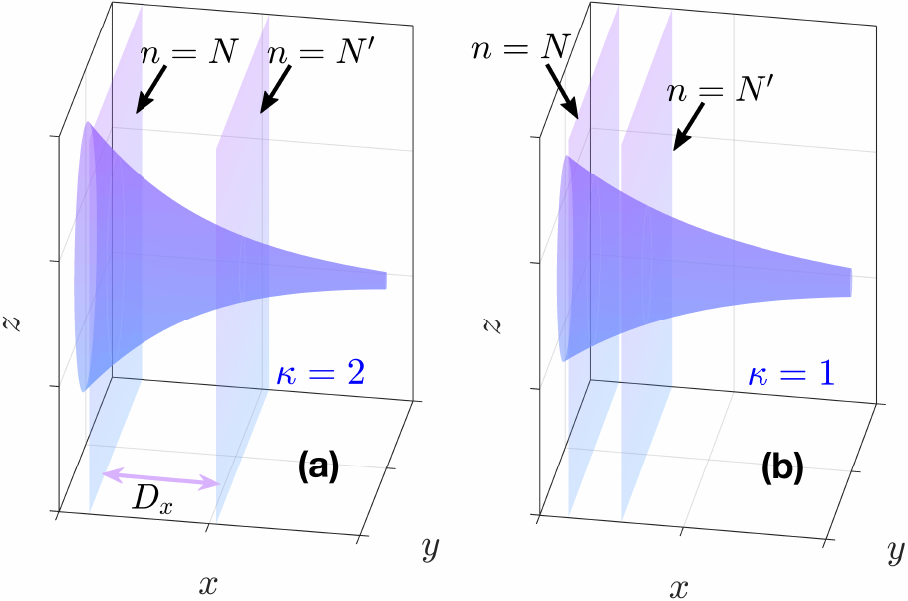}
\caption{Funnel-shaped hyperbolic surfaces for curvatures (a) $\kappa=2$ and (b) $\kappa=1$. The semitransparent planes illustrate the locations of the two emitter-lattice coupling points (i.e., $n=N$ and $n=N'$) in the effective curved space.}
\label{Fcurve}
\end{figure}


Very recently, it has been shown that a class of non-Hermitian systems can simulate quantum systems in \emph{curved spaces}~\cite{curveLv,DualityPRB}. In particular, an HN model can be mapped to a two-dimensional hyperbolic lattice whose curvature is determined by the non-Hermiticity $\gamma$. While conventional schemes of implementing hyperbolic lattices require that the spatial structure of a physical system has to be distorted to become curved~\cite{Ccurve1,Ccurve2,Ccurve3,Ccurve4}, this duality (between the non-Hermitian systems and their hyperbolic counterparts) allows for simulating hyperbolic spaces using lattices with fixed lattice constants. For the HN model considered in this work, it is equivalent to a hyperbolic lattice with curvature $-\kappa=-4\ln^{2}(|\beta|)$. The coordinates of lattice sites in the effective curved space are given by $x_{n}=x_{0}\exp(n\sqrt{\kappa})$ ($x_{0}$ is the coordinate of the $0$th lattice site of the HN model)~\cite{curveLv}.

Such a duality can be visualized using the hyperbolic surface on the Poincar\'{e} half-plane (i.e., pseudosphere)~\cite{curveLv,Bulletin}, as shown in Fig.~\ref{Fcurve}. The hyperbolic surface has a funnel shape, with the variation in the cross-section circumference along the $x$-axis determined by the curvature $-\kappa$. More specifically, the unit vectors $\vec{u}$, $\vec{v}$, and $\vec{w}$ of the three spatial dimensions (in the effective curved space) satisfy
\begin{equation}
\mleft[u-\frac{\text{arcsech}(\sqrt{(v^{2}+w^{2})\kappa})}{\sqrt{\kappa}}\mright]^{2}+v^{2}+w^{2}=\frac{1}{\kappa}.
\label{surface}
\end{equation}
We also illustrate in Fig.~\ref{Fcurve} the two emitter-lattice coupling points (the semitransparent planes). It is clear that the separation $D_{x}$ between the coupling points in the effective curved space increases with $\kappa$ according to $D_{x}=x_{0}[\exp(N'\sqrt{\kappa})-\exp(N\sqrt{\kappa})]$. That is to say, both the strengths and separations of the coupling points of a giant emitter have to be tailored (according to the lattice curvature) in order to observe conventional giant-emitter interference effects in such a curved space. This provides a guidance for studying and implementing giant emitters in curved spaces.

\end{document}